\documentclass[review,authoryear,3p,times,12pt]{elsarticle}

\usepackage{multirow,setspace,times,amssymb,amsmath,graphicx,color,rotating,subfigure,url}
\usepackage{lineno,color}
\usepackage{natbib}
\usepackage{booktabs}%
\usepackage{longtable}%
\graphicspath{{Figures/}}
\usepackage[table]{xcolor}
\usepackage{tabularx}
\usepackage{graphicx} 
\usepackage{mathrsfs}
\usepackage{makecell}
\usepackage{threeparttable}
\usepackage[bookmarks=true,colorlinks,linkcolor=blue,anchorcolor=blue,citecolor=blue,unicode]{hyperref}
\usepackage{bookmark}
\usepackage{booktabs}
\usepackage{array,caption,threeparttable}
\usepackage[font=small,labelfont=bf,labelsep=none]{caption}
\usepackage[title]{appendix}
\usepackage{amsmath}

\hypersetup{CJKbookmarks=true}%
\journal{International Review of Economics \& Finance }

\usepackage{caption}
\begin{document}

\begin{frontmatter}

\title{The impact of external uncertainties on the extreme return connectedness between food, fossil energy, and clean energy markets}

\author[HNUST]{Ting Zhang}
\ead{tzhang@hnust.edu.cn}
\author[SB,RCE]{Hai-Chuan Xu\corref{CorAuth}}
\ead{hcxu@ecust.edu.cn}
\author[SB,RCE,DM]{Wei-Xing Zhou\corref{CorAuth}}
\ead{wxzhou@ecust.edu.cn}
\cortext[CorAuth]{Corresponding authors. Corresponding to: 130 Meilong Road, P.O. Box 114, School of Business, East China University of Science and Technology, Shanghai 200237, China.}

\address[HNUST]{School of Business, Hunan University of Science and Technology, Xiangtan 411201, China}
\address[SB]{School of Business, East China University of Science and Technology, Shanghai 200237, China}
\address[RCE]{Research Center for Econophysics, East China University of Science and Technology, Shanghai 200237, China}
\address[DM]{School of Mathematics, East China University of Science and Technology, Shanghai 200237, China}

\begin{abstract}
We investigate the extreme return connectedness between the food, fossil energy, and clean energy markets using the quantile connectedness approach, which combines the traditional spillover index with quantile regression. Our results show that return connectedness at the tails (57.91\% for the right tail and 61.47\% for the left tail) is significantly higher than at the median (23.02\%). Furthermore, dynamic analysis reveals that connectedness fluctuates over time, with notable increases during extreme events. Among these markets, fossil energy market consistently acts as the net receiver, while clean energy market primarily serves as the net transmitter. Additionally, we use linear and nonlinear ARDL models to examine the role of external uncertainties on return connectedness. We find that climate policy uncertainty (CPU), geopolitical risk (GPR), and the COVID-19 pandemic significantly impact median connectedness, while economic policy uncertainty (EPU), GPR, and trade policy uncertainty (TPU) are crucial drivers of extreme connectedness. Our findings provide valuable insights for investors and policymakers on risk spillover effects between food and energy markets under both normal and extreme market conditions.
\end{abstract}

\begin{keyword}
 Extreme spillover \sep Energy market \sep Food market \sep External uncertainty
\\
  JEL: G28, C1, Q4, Q1
\end{keyword}

\end{frontmatter}


\section{Introduction}
\label{S1:Introduction}

Energy and food security are essential for sustainable development and human well-being \citep{TaghizadehHesary-Rasoulinezhad-Yoshino-2019-EnergyPolicy,Guo-Tanaka-2022-EnergyEcon}. With increasing mechanization in agriculture, energy (particularly fossil fuels, like crude oil, coal, and their derivatives) has become a key input in agricultural production, affecting irrigation, transportation, and the production of chemicals and fertilizers \citep{Zimmer-Marques-2021-Energy}. Given that energy costs represents a large proportion of production expenses, food prices experienced sharp increases during the periods of energy crisis \citep{Youssef-Mokni-2021-IntEconJ}. The connection between energy and food prices is commonly mediated through production costs, a relationship demonstrated by numerous studies \citep{Ericsson-Rosenqvist-Nilsson-2009-BiomassBioenerg,Georgiou-Acha-Shah-Markides-2018-JCleanProd}. The explosion of global biofuel industry in the second half of 2000s provides a new dimension to this connection, with rising energy prices triggering demand for biofuels made from crops like corn and soybeans \citep{Myers-Johnson-Helmar-Baumes-2014-AmJAgrEcon,Yoon-2022-RenewEnergy,Tanaka-Guo-Wang-2023-GCBBioenergy}. This, in turn, leads to higher food prices as corn and soybeans compete with other crops for land, water, and profits, and then raise the production costs of other food, like oil, meat and dairy \citep{Atems-Mette-2024-JEnvironPlanManag}. In addition, broader macroeconomic factors, including inflation and economic policy, contribute to the co-movement of food and energy prices \citep{Adil-Bhatti-Waqar-Amin-2022-JPublicAff}.

In recent years, climate change has emerged as a global challenge faced by human-beings. Scholars believe that climate change is mainly caused by human activities, among which agricultural production is a typical sector with high energy consumption and high pollution \citep{Hartter-Hamilton-Boag-Stevens-Ducey-Christoffersen-Oester-Palace-2018-ClimServ}. In this context, many governments are embarking on a green transformation of agricultural production. For example, Indian central government has pledged to provide solar power to farms as part of efforts to reduce reliance of agriculture on conventional energy sources \citep{Chatterjee-2024-EnergyPolicy}. The availability and cost of clean energy are increasingly recognized as critical for ensuring food security, with studies highlighting both positive and negative influences of clean energy on agricultural outcomes \citep{Haque-Khan-2022-JAgribusDevEmergEcon,Li-Agene-Gu-Osabohien-Jaaffar-2024-FrontSustainFoodSyst}. Furthermore, \cite{Han-Zhang-Li-2022-ApplEnergy} take the case of China to study the rural energy transition in developing countries. Their results reveal that urbanization has a positive effect on the usage of clean energy in agriculture. With the energy transformation of the whole society and the development of green agriculture, prices of clean energy and food are increasingly linked.

Research on food-energy nexus is growing \citep{Abdelradi-Serra-2015-EnergyEcon,Lucotte-2016-EconLett,Diab-Karaki-2023-EnergyEcon}. On the one hand, since food and energy are essential substances for the development of human society, the interaction between food and energy prices has a noteworthy effect on economic stability \citep{Lucotte-2016-EconLett}. On the other hand, the financialization of commodities has made food and energy commodities important asset classes for global investors, increasing the vulnerability of commodity prices to financial market factors, such as exchange rate \citep{Adil-Bhatti-Waqar-Amin-2022-JPublicAff}. Indeed, researchers find that food and energy commodity markets have negative relations with equity markets, and they have positive relations themselves \citep{Han-Zhou-Yin-2015-EnergyEcon}. Therefore, the dependencies of food and energy prices provide important references for investors in portfolio hedging and risk management.

Existing works mainly focus on food-oil nexus and have came to mixed results. Most of these research find a significantly positive relationship between food and oil prices \citep{Mohammed-2022-OPECEnergyRev,Yu-Peng-Zakaria-Mahmood-Khalid-2023-JBusEconManag}, while others find weak linkages \citep{Zmami-BenSalha-2019-Economies} or heterogeneous and asymmetric impacts across different food categories \citep{Chen-Yan-Kang-2022-RevDevEcon}. In addition, there are also studies that examine the relationship between food and other fossil energy, such as coal, natural gas, and gasoline \citep{Diab-Karaki-2023-EnergyEcon,Vatsa-Miljkovic-Baek-2023-JAgricEcon,Miljkovic-Vatsa-2023-IntRevFinancAnal}. However, although the energy applied in the agricultural sector is transmitting from fossil energy to clean energy gradually and this process will continue to move forward, there remains a lack of studies that incorporate clean energy into the research framework \citep{Chatterjee-2024-EnergyPolicy,Haque-Khan-2022-JAgribusDevEmergEcon}. 

In this work, we focus on the interactions between food, fossil energy, and clean energy markets. To begin with, we analysis their return connectedness at normal and extreme market conditions by using a quantile-based spillover approach which combines the \cite{Diebold-Yilmaz-2012-IntJForecast} spillover index with a quantile regression approach. Our results show that the connectedness between food, fossil energy, and clean energy markets are much stronger at both the extreme upper and lower quantiles than at the conditional median. Moreover, the return connectedness is asymmetric, specifically, it is higher at the left tail than at the right tail. We next conduct the dynamic analysis using the rolling window method to capture the time-varying characteristics of connectedness. The results reveal that the total spillover fluctuates significantly during the sample periods, and it increases notably when extreme events occur, such as the signing and implementation of the Paris Agreement in 2015 and 2016, the withdrawal of the US from the Paris Agreement in 2017 and its return in early 2021, the COVID-19 pandemic in 2020, and the Russia-Ukraine conflict in 2022. The index at tails is less volatile than at the median. In addition, the net spillover analysis indicates that fossil energy market always act as the net receiver, while clean energy market plays more role of a net transmitter. This is consistent with the results of previous research that clean energy has a significant spillover effect to fossil energy as energy consumption transmitting from fossil fuels to clean energy \citep{Raza-Khan-Benkraiem-Guesmi-2024-IntRevFinancAnal}.

Motivated by the aforementioned results of the fluctuations of connectedness, we consider the effects of external uncertainties, including economic policy uncertainty \citep{Baker-Bloom-Davis-2016-QJEcon}, climate policy uncertainty \citep{Gavriilidis-2021-SSRN}, trade policy uncertainty \citep{Baker-Bloom-Davis-2016-QJEcon}, and geopolitical risk index \citep{Caldara-Iacoviello-2022-AmEconRev}. Besides, we take COVID-19 as a dummy variable, which takes the value of 1 during the pandemic between January 2020 and December 2020, and 0 otherwise. We apply the linear and nonlinear autoregressive distributed lags (ARDL) models, incorporating the logarithm of these uncertainty indexes as the predictor variables and the total return connectedness index as the dependent variable. We run the regression for the connectedness at the conditional median and the extreme quantiles. For the median connectedness, CPU, GPR, and COVID-19 pandemic has significant impact. For extreme connectedness, EPU, GPR, and TPU are key drivers. Addityonally, the results of NARDL models reveal the asymmetric effects of external uncertainties, specifically, CPU has a short-term asymmetric effect on connectedness at the extreme upper quantile ($\tau$ = 0.95), while for the long-term asymmetry, EPU is significant on the conditional median, and CPU, TPU, and GPR are significant on the extreme lower quantile ($\tau$ = 0.05).

The remainder of the paper is organized as follows. Section~\ref{S2:Literature} reviews the related studies. Section~\ref{S3:Methodology:Data} introduces the quantile connectedness methods and describes the data we use. Section~\ref{S4:EmpAnal} provides the empirical results regarding the spillovers between food, fossil energy, and clean energy markets under normal and extreme conditions. Section~\ref{S5:External:Uncertainties} explores the impacts of external uncertainties on the connectedness between these markets, and Section~\ref{S6:Conclude} concludes the work and presents policy implications.


\section{Literature Review}
\label{S2:Literature}

The relationship between food and energy markets has been the focus of growing literature. From the spillover effects point of view, existing studies consider the price level \citep{Youssef-Mokni-2021-IntEconJ} and volatility level transmissions \citep{Chatziantoniou-Degiannakis-Filis-Lloyd-2021-EnergyJ}. From the perspective of methodologies, the literature includes linear \citep{Roman-Gorecka-Domagala-2020-Energies} and nonlinear methods \citep{Yu-Peng-Zakaria-Mahmood-Khalid-2023-JBusEconManag}.

 Early literature is more undertaken the linear framework to study the relationship between food and energy markets \citep{Hassouneh-Serra-Goodwin-Gil-2012-EnergyEcon,Roman-Gorecka-Domagala-2020-Energies}. \cite{Hassouneh-Serra-Goodwin-Gil-2012-EnergyEcon} find a long-run equilibrium relationship between agriculture and crude oil prices by using multivariate linear regression method, and they confirm the biofuel channel as the effect mechanism. \cite{Roman-Gorecka-Domagala-2020-Energies} employ the cointegration test and Granger causality test to examine the linkage between crude oil and the price indexes of five categories of food. Their findings reveal a long-term relationship between crude oil and meat prices, while shorter-term linkages are observed between crude oil and cereal or oil prices. Additionally, \cite{Fasanya-Akinbowale-2019-Energy} provide the evidence of the interdependence between crude oil and food prices from the perspective of spillovers by using the \cite{Diebold-Yilmaz-2012-IntJForecast} method. 
 
In more recent studies, researchers explore the non-linear characteristics of this relationship. They find that the interaction between food and energy prices exhibits diverse features at different market conditions \citep{Youssef-Mokni-2021-IntEconJ}. \cite{Youssef-Mokni-2021-IntEconJ} apply the MRS-QR model to examine how food prices respond to different oil price shocks. Their results confirm that the contagion effect between the two markets during the periods of crisis, but the reaction of food price to oil price shocks changes with the structure of the shocks. \cite{Yu-Peng-Zakaria-Mahmood-Khalid-2023-JBusEconManag} use the quantile-on-quantile estimation method and find that oil and food prices present nonlinear dependences, specifically, the correlation is negative for lower and medium quantiles and positive for higher quantile. Along with \cite{Yu-Peng-Zakaria-Mahmood-Khalid-2023-JBusEconManag}, \cite{Sun-Gao-Raza-Shah-Sharif-2023-Energy} also employ the quantile-on-quantile method and categorize oil prices into demand and supply shocks. According to their results, food price subindices are correlated with oil price shocks in varying degrees, depending on the quantile and the type of shock. Similarly, \cite{Wang-Chavas-Li-2024-AgricEcon} adopt the quantile impulse response approach and find that the speed of food prices adjust to oil price shocks differs across quantiles. \cite{Hanif-Hernandez-Shahzad-Yoon-2021-QRevEconFinanc} focus on tail dependence and reveal that oil and food prices are independent at both the left and right tails. Nonlinear autoregressive distributed lags (NARDL) models are also widely used in examining their nonlinear relationship. \cite{Almalki-Hassan-BinAmin-2022-ApplEcon} and \cite{Chowdhury-Meo-Uddin-Haque-2021-Energy} adopt this method and confirm the asymmetric effects of energy pricesss on food prices.

Although most existing studies focus on the relationships on price level, several works have also explored the volatility spillover effect between food and energy markets \citep{Chatziantoniou-Degiannakis-Filis-Lloyd-2021-EnergyJ}. \cite{Chatziantoniou-Degiannakis-Filis-Lloyd-2021-EnergyJ} employ the HAR and MIDAS-HAR approach to examine the out-of-sample predictability of oil price volatility on food price volatility. Their findings indicate that oil price volatility has weak effect in the out-of-sample prediction of food price volatility, which contrasts with the in-sample results from previous studies \citep{Algieri-Leccadito-2017-EnergyEcon,Zhang-Broadstock-2020-IntRevFinancAnal}. \cite{Ucak-Yelgen-Ari-2022-Bio-basedApplEcon} use the \cite{Diebold-Yilmaz-2012-IntJForecast} approach to examine the volatility spillover between energy and foods markets, revealing that the volatility spillover is significant from energy prices to vegetable prices but not to fruit prices. \cite{Liu-Serletis-2024-ApplEcon} adopt the GARCH-in-Mean copula models to study the volatility dynamics and dependence of crude oil and major agricultural commodity prices. 

Besides crude oil, the most extensively studied energy category, researchers have also explored the relationship between food and other types of energy. For example, \cite{Miljkovic-Vatsa-2023-IntRevFinancAnal} adopt the dynamic time warping method and find the lead-lag relationship between coal, natural gas prices and six major agricultural commodities. \cite{Vatsa-Miljkovic-Baek-2023-JAgricEcon} explore the linkages between natural gas and cereals. The authors find that cereal prices respond to natural gas price shocks with slight and transitory characteristics. Moreover, gasoline, as one of the most important derivatives of crude oil, its price shocks also have a significantly positive effect on food prices \citep{Diab-Karaki-2023-EnergyEcon}.

Our work contributes to the literature on the return connectedness between food, fossil energy, and clean energy markets, and the impacts of external uncertainties. In a similar study, \cite{Yousfi-Bouzgarrou-2024-EnvironSciPollutRes} examine the volatility connectedness between these markets by employing the DCC-GARCH approach, and further analysis the effect of EPU and GPR on the connectedness using the quantile-on-quantile model. Their findings reveal that the dynamic volatility spillovers between these markets are sensitive to EPU and GPR. There are three key differences that distinguish our work from theirs. First, while \cite{Yousfi-Bouzgarrou-2024-EnvironSciPollutRes} focus on volatility connectedness, we center our analysis on the price level. Second, their study uses sub-indices of fossil and clean energy, while we opt for a more comprehensive index as the representative measure for both fossil and clean energy. Lastly, \cite{Yousfi-Bouzgarrou-2024-EnvironSciPollutRes} use the quantile-on-quantile model to analysis the effects of individual uncertainty separately. This method focuses on the impact of single factor and is not suitable for the multivariate case. In contrast, we adopt the linear and nonlinear ARDL model to capture the effects of multiple uncertainties, thus offering a more holistic understanding of the role of external uncertainties.

\section{Methodology and Data}
\label{S3:Methodology:Data}

\subsection{Quantile TVP-VAR-DY approach}
\label{S3.1:TVP-VAR-DY}

Following \cite{Koenker-Bassett-1976-Econometrica}, for different quantiles $\tau \in (0,1)$, the dependence of $y_t$ on $x_t$ can be estimated using the following equation:
\begin{equation}
\label{Eq:y_t}
    y_t=c(\tau)+\sum^p_{i=1}B_i(\tau)y_{t-i}+e_t(\tau), t=1,\dots , T
\end{equation}

According to \cite{Koop-Pesaran-Potter-1996-JE} and \cite{Pesaran-Shinr-1998-EL}, the generalized forecast error variance decomposition (GFEVD) with forecast horizon $H$ is calculated as follows:
\begin{equation}
    \label{GFEVD}
    Q^g_{ij}(H)=\frac{\sigma^{-1}_{jj}\sum^{H-1}_{h=0}(e^{'}_ih_h\sum e_j)^2}{\sum^{H-1}_{h=0}(e^{'}_ih_h\sum e_j)},
\end{equation}

The normalization of each vector in the decomposition matrix is:
\begin{equation}
    \label{Eq:Normalization}
    \widetilde{Q}^g_{ij}(H)=\frac{Q^g_{ij}(H)}{\sum^N_{j=1}Q^g_{ij}(H)}.
\end{equation}

Various quantile spillover measures can be defined using the GFEVD method based on the approach of \cite{Diebold-Yilmaz-2012-IntJForecast}:
\begin{equation}
    \label{Eq:TSI}
    TSI(\tau)=\frac{\sum^N_{i=1}\sum^N_{j=1, i\neq j}\omega^h_{ij}(\tau)}{\sum^N_{i=1}\sum^N_{j=1}\omega^h_{ij}(\tau)} \times 100.
\end{equation}
 \begin{equation}
    \label{Eq:DSI1}
    S_{\text{all}\rightarrow i}(\tau)=\frac{\sum^N_{j=1, i\neq j}\omega^h_{ij}(\tau)}{\sum^N_{j=1}\omega^h_{ij}(\tau)} \times 100
 \end{equation}
 \begin{equation}
    \label{Eq:DSI2}
    S_{i \rightarrow \text{all}}(\tau)=\frac{\sum^N_{j=1, i\neq j}\omega^h_{ji}(\tau)}{\sum^N_{j=1}\omega^h_{ji}(\tau)} \times 100
 \end{equation}
\begin{equation}
    \label{Eq:NSI}
    NS_i(\tau)=S_{i \rightarrow \text{all}}(\tau)-S_{\text{all}\rightarrow i}(\tau)
\end{equation}
$TSI$ indicates the total spillover index. $S_{\text{all}\rightarrow i}$ and $S_{i \rightarrow \text{all}}$ represent the directional spillover index of index $i$ received from all indices and transfer to all indices, respectively. $NS_i$ is the net spillover index that can be calculated by the disparity between $S_{\text{all}\rightarrow i}(\tau)$ and $S_{i \rightarrow \text{all}}(\tau)$, wherein a positive (negative) value indicates the net spillover transmitter (recipient).

\subsection{Autoregressive Distributed Lag (ARDL) model}
\label{S3.2:ARDL}

In order to test the long-run and short-run effects of uncertainties on the spillovers, we consider the Autoregressive Distributed Lag (ARDL) model proposed by \citep{Pesaran-Shin-Smith-2001-JAEm} as follows:
\begin{multline}
    \Delta ln{TSI_t} = \alpha_0 + \alpha_1 ln{TSI_{t-1}} + \alpha_2 ln EPU_{t-1} + \alpha_3 ln CPU_{t-1} + \alpha_4 ln TPU_{t-1} + \alpha_5 ln GPR_{t-1} + \alpha_6 COVID-19 \\
    + \sum^{n_1}_{i=1} \beta_i \Delta ln TSI_{t-i} + \sum^{n_2}_{i=0} \gamma_i \Delta ln EPU_{t-i} + \sum^{n_3}_{i=0} \lambda_i \Delta ln CPU_{t-i} + \sum^{n_4}_{i=0} \delta_i \Delta ln TPU_{t-i} + \sum^{n_5}_{i=0} \omega_i \Delta ln GPR_{t-i} + \epsilon_t
    \label{Eq: ARDL}
\end{multline}
where $\Delta$ is the first different operator, $n_i$ $(i = 1,2, \dots 5)$ is the optimal lag order determined by the Akaike information criterion (AIC), and $\epsilon_t$ refers to the error term.

The existence of long-run cointegration can be examined by using the bound test \citep{Pesaran-Shin-Smith-2001-JAEm}. The null hypothesis of no cointegration among underlying variables is $H_0: \alpha_1 = \alpha_2 = \alpha_3 = \alpha_4 = \alpha_5 = \alpha_6 = 0$ against the alternative hypothesis $H_1: \alpha_1 \neq \alpha_2 \neq \alpha_3 \neq \alpha_4 \neq \alpha_5 \neq \alpha_6 \neq 0$. If the long-run cointegration exists, then we can construct an error correction term (ECT) and model~(\ref{Eq: ARDL}) can be converted to:
\begin{align}
    \Delta ln{TSI_t} &= \alpha_0 + \sum^{n_1}_{i=1} \beta_i \Delta ln TSI_{t-i} + \sum^{n_2}_{i=0} \gamma_i \Delta ln EPU_{t-i} + \sum^{n_3}_{i=0} \lambda_i \Delta ln CPU_{t-i} \notag \\
    &+ \sum^{n_4}_{i=0} \delta_i \Delta ln TPU_{t-i} + \sum^{n_5}_{i=0} \omega_i \Delta ln GPR_{t-i} + \phi ECT_{t-1} + \epsilon_t
    \label{Eq: ARDL-ECM}
\end{align}

Furthermore, we construct nonlinear autoregressive distributed lag (NARDL) model of \cite{Shin-Yu-Greenwood-Nimmo-2014-Festschrift}. In NARDL model, the exogenous variables are decomposed into positive and negative partial sum series to capture the asymmetric relationships between total spillovers and the external uncertainties:
\begin{align}
    X_t^+ = \sum^t_{j=1} \Delta X_j^+ = \sum^t_{j=1} \max(\Delta X_j, 0) \\
    X_t^- = \sum^t_{j=1} \Delta X_j^- = \sum^t_{j=1} \min(\Delta X_j, 0)
\end{align}
where $X$ refer to the external uncertainty index. Then, we compute the decomposition of $ln EPU$, $ln CPU$, $ln TPU$, and $ln GPR$ and represent them into the NARDL model as follows:
\begin{align}
 \Delta ln{TSI_t} &= \alpha_0 + \alpha_1 ln{TSI_{t-1}} + \sum^{n_1}_{i=1} \beta_i \Delta ln TSI_{t-i} \notag \\
 &+ \alpha_2^+ ln EPU_{t-1}^+ + \alpha_2^- ln EPU_{t-1}^- + \sum^{n_2}_{i=0} (\gamma_i^+ \Delta ln EPU_{t-i}^+ + \gamma_i^- \Delta ln EPU_{t-i}^-) \notag \\
 &+ \alpha_3^+ ln CPU_{t-1}^+ + \alpha_3^- ln CPU_{t-1}^- + \sum^{n_3}_{i=0} (\lambda_i^+ \Delta ln CPU_{t-i}^+ + \lambda_i^- \Delta ln CPU_{t-i}^-) \notag \\
 &+ \alpha_4^+ ln TPU_{t-1}^+ + \alpha_4^- ln TPU_{t-1}^- + \sum^{n_4}_{i=0} (\delta_i^+ \Delta ln TPU_{t-i}^+ + \delta_i^- \Delta ln TPU_{t-i}^-) \notag \\
 &+ \alpha_5^+ ln GPR_{t-1}^+ + \alpha_5^- ln GPR_{t-1}^- + \sum^{n_5}_{i=0} (\omega_i^+ \Delta ln GPR_{t-i}^+ + \omega_i^- \Delta ln GPR_{t-i}^-) \notag \\
 &+ \alpha_6 COVID-19 + \epsilon_t
    \label{Eq: NARDL}
\end{align}

Accordingly, the ECT form NARDL model can be written as:
\begin{align}
 \Delta ln{TSI_t} &= \alpha_0 + \sum^{n_1}_{i=1} \beta_i \Delta ln TSI_{t-i}  + \sum^{n_2}_{i=0} (\gamma_i^+ \Delta ln EPU_{t-i}^+ + \gamma_i^- \Delta ln EPU_{t-i}^-) \notag \\
 &+ \sum^{n_3}_{i=0} (\lambda_i^+ \Delta ln CPU_{t-i}^+ + \lambda_i^- \Delta ln CPU_{t-i}^-)  + \sum^{n_4}_{i=0} (\delta_i^+ \Delta ln TPU_{t-i}^+ + \delta_i^- \Delta ln TPU_{t-i}^-) \notag \\
 &+ \sum^{n_5}_{i=0} (\omega_i^+ \Delta ln GPR_{t-i}^+ + \omega_i^- \Delta ln GPR_{t-i}^-)  + \phi ECT_{t-1} + \epsilon_t
    \label{Eq: NARDL-ECM}
\end{align}

If $\alpha_i^+ \neq \alpha_i^-$ $(i =2,3, \dots, 5)$, we would conclude that the effect is asymmetric in the long-run. Similarly, if $\gamma_i^+ \neq \gamma_i^-$, $\lambda_i^+ \neq \lambda_i^-$, $\delta_i^+ \neq \delta_i^-$, or $\omega_i^+ \neq \omega_i^-$, then the asymmetric effect exists for the corresponding variable in the short-run. We also examine the long-run cointegration by using the bound test \citep{Pesaran-Shin-Smith-2001-JAEm}.

\subsection{Data description}
\label{S3.3:Data}

\begin{figure}[!ht]
\centering
\includegraphics[width=0.325\textwidth]{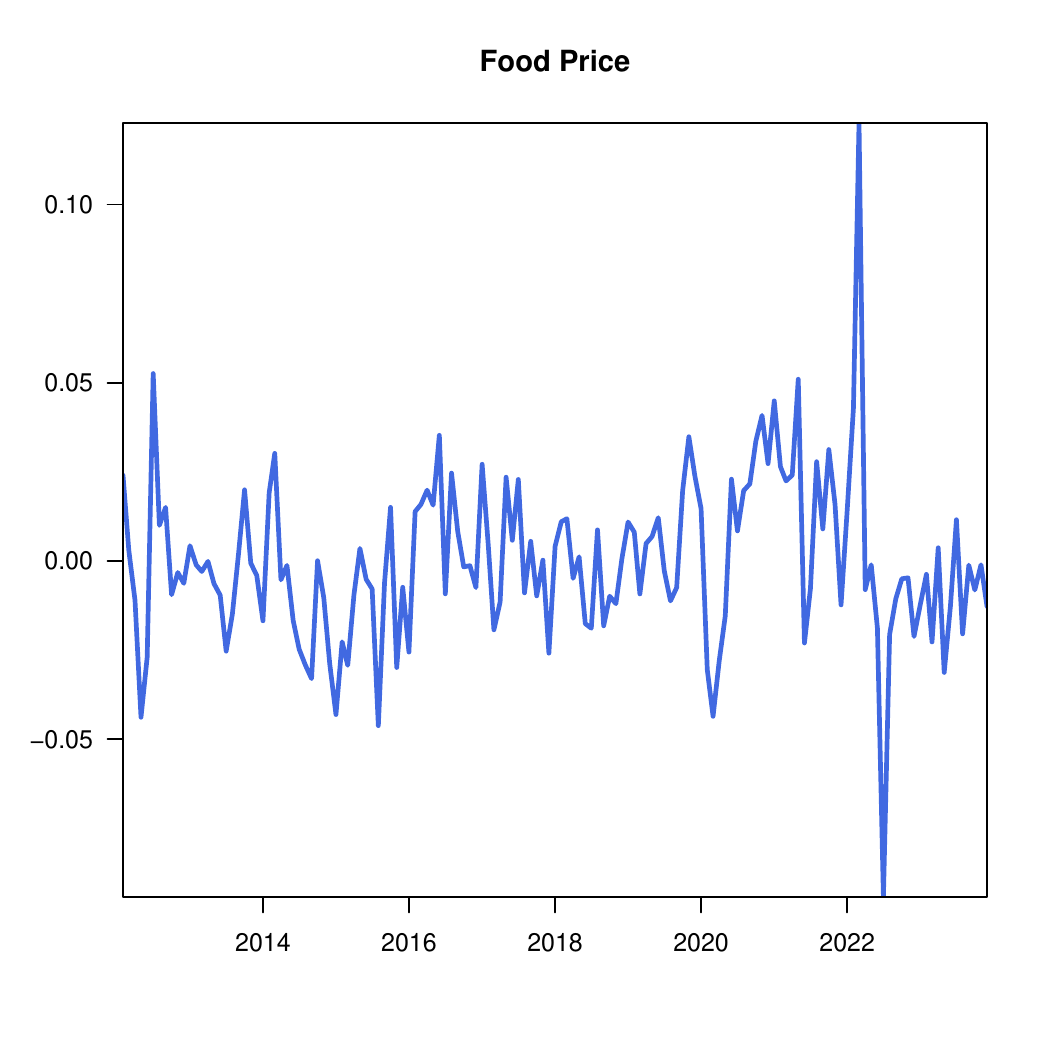}
\includegraphics[width=0.325\textwidth]{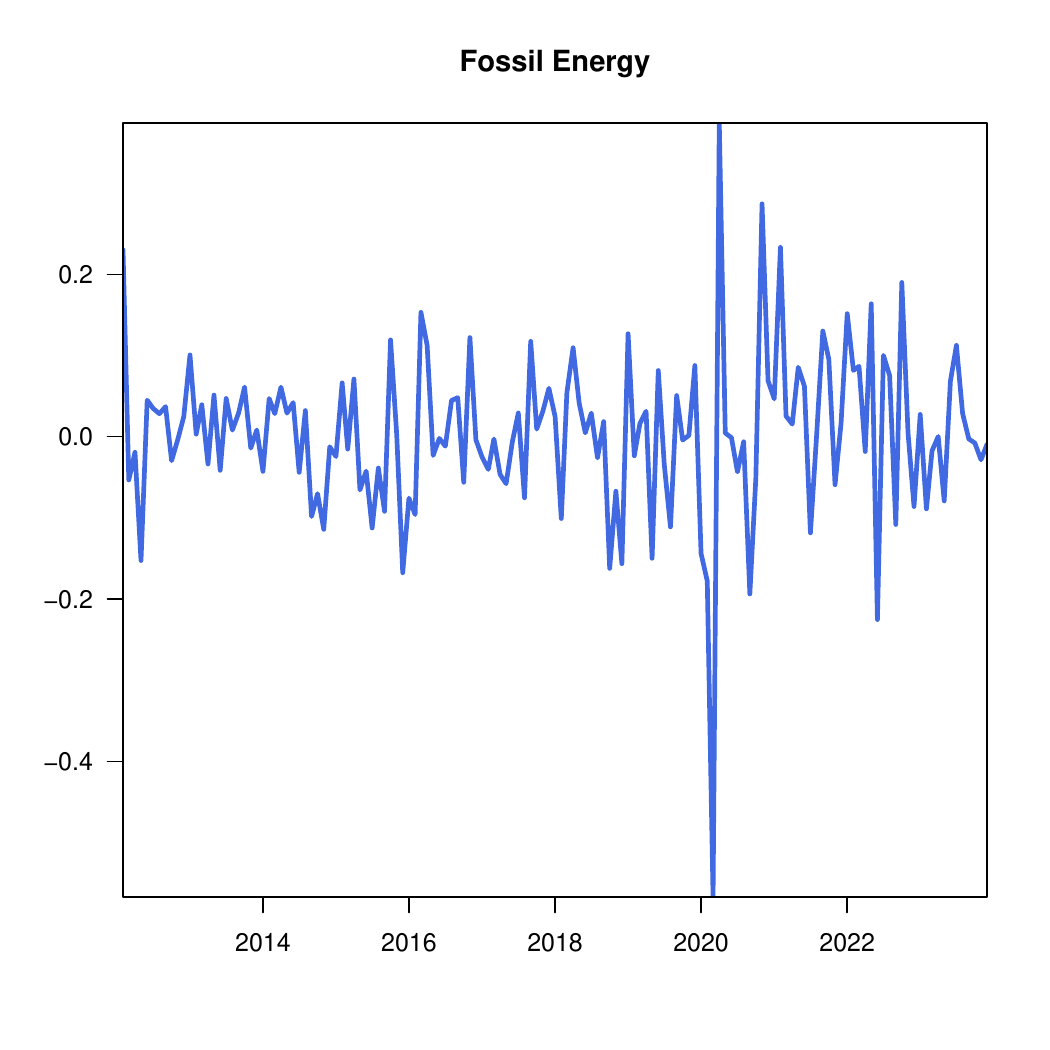}
\includegraphics[width=0.325\textwidth]{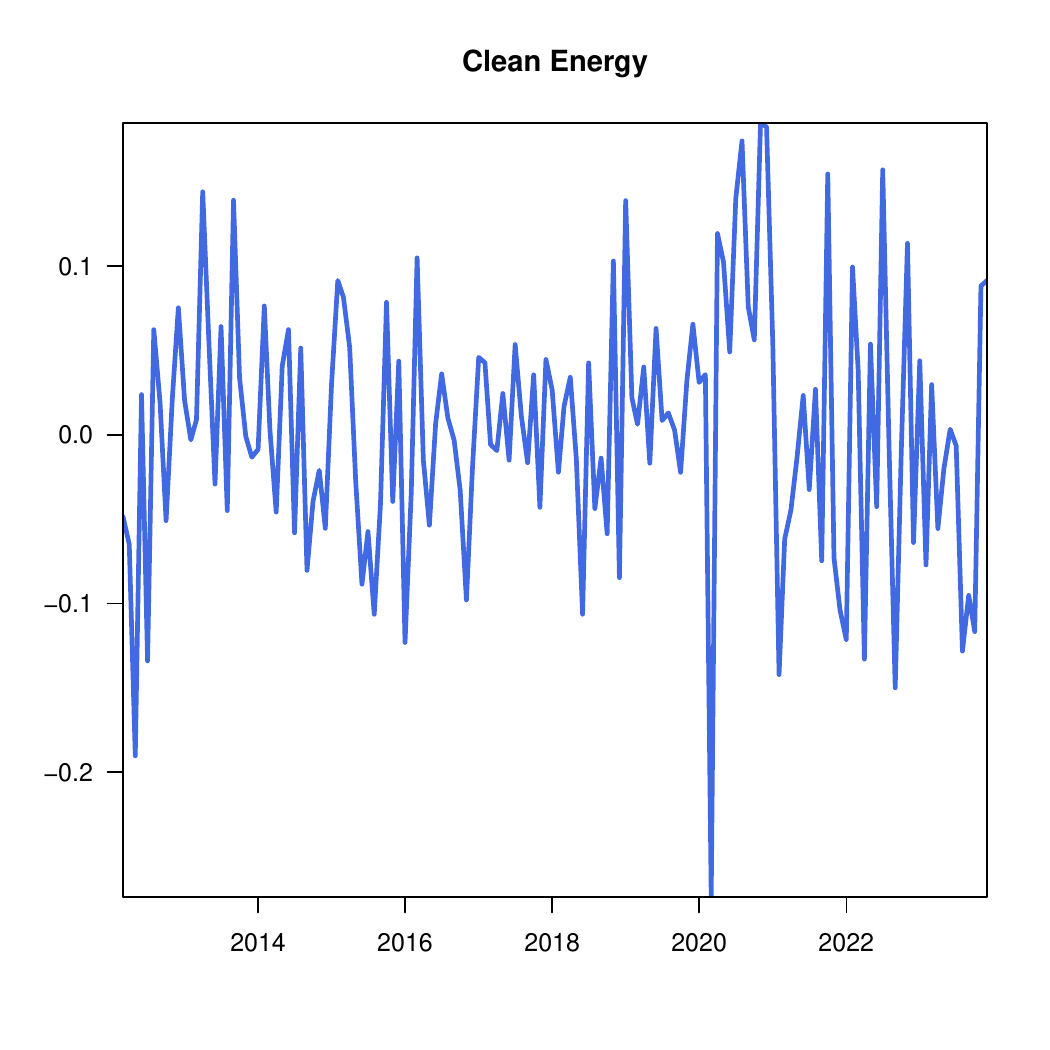}
 \vspace{-10bp}
\caption{Retruns of food price index, fossil energy, and clean energy ETFs.}
\label{Fig:Return}
\end{figure}

To track the price changes in the food market, we adopt the Food Price Index (FPI) released by the Food and Agricultural Organization (FAO).\footnote{We obtain FPI from \url{https://www.fao.org/}.} For the fossil energy market,
we use the iShares US Oil \& Gas Exploration \& Production ETF (IEO), which tracks US-based companies involved in the exploration and production of oil and gas. For the clean energy market, we use the iShares Global Clean Energy ETF (ICLN), which tracks companies involved in the production of renewable energy sources like solar and wind.\footnote{Data on these ETFs is extracted from the Wind Database (\url{https://www.wind.com.cn/}).} The sample period spans from January 2012 to December 2023, and all data are at a monthly frequency. Fig.~\ref{Fig:Return} depicts the evolution of the monthly log returns of the FPI, IEO, and ICLN ETFs. Notably, the food price index peaked in early 2022 due to the Russia-Ukraine conflict, followed by a sharp decline. Both fossil energy and clean energy ETFs experienced a rapid drop in early 2020 due to the COVID-19 pandemic, followed by a subsequent rebound.

Panel A of Table~\ref{tab:statistics} reports the descriptive statistics and results of the unit root test for the log returns of the variables. The mean returns are negative for food price and clean energy, and positive for fossil energy. The fossil energy market exhibits the highest volatility, with a standard deviation of 0.1020, followed by clean energy at 0.0898, both exceeding the food market's standard deviation of 0.0238. The food price index is positively skewed, while both fossil energy and clean energy are negatively skewed. The kurtosis of all variables are larger than three, indicating the thick tails of the distributions. Moreover, Jarque-Bera's statistics show that the variables are not normally distributed. The ADF test results indicate that all variables are stationary. Panel B of Table~\ref{tab:statistics} presents the Pearson correlation coefficients, highlighting a significantly positive correlation between the food and fossil energy market. However, the correlation is not significant between food and clean energy markets. Fossil energy and clean energy markets also exhibit positive correlation at 0.231.

\begin{table*}[!ht]
  \centering
  \setlength{\abovecaptionskip}{0.1cm}
  \caption{Descriptive statistics.}
     \setlength{\tabcolsep}{9pt}
    \begin{threeparttable}
    \begin{tabular}{cccccccc}
    \toprule
    Variable & Mean & Std. Dev. & Skewness & Kurtosis & Jarque-Bera & ADF\\
    \midrule
    \multicolumn{7}{l}{\it{Panel A: Descriptive statistics}}\\
    Food Price & -0.0002 & 0.0238 & $~~~$0.6564 & $~~$8.2797 & $176.3627^{***}$ &  $-3.6251^{**~}$\\
    Fossil Energy & $~$0.0037 & 0.1020 & $-0.7542$ & 10.1492 & $318.0887^{***}$ &  $-4.7320^{***}$\\
    Clean Energy & -0.0005 & 0.0898 & $-1.7873$ & 12.9431 & $665.1999^{***}$ & $-4.9140^{***}$\\
    \midrule
    & \multicolumn{2}{c}{Food Price} & \multicolumn{2}{c}{Fossil Energy} & \multicolumn{2}{c}{Clean Energy}\\
    \midrule
    \multicolumn{7}{l}{\it{Panel B: Correlations}}\\
    Food Price  & \multicolumn{2}{c}{1.000$~~~$} &  \multicolumn{2}{c}{} & \multicolumn{2}{c}{} \\
    Fossil Energy & \multicolumn{2}{c}{$0.263^{***}$} &\multicolumn{2}{c}{ 1.000$~~~$} & \multicolumn{2}{c}{}  \\
    Clean Energy  & \multicolumn{2}{c}{0.107$~~~$} & \multicolumn{2}{c}{$0.231^{***}$} & \multicolumn{2}{c}{1.000} \\
    \bottomrule
    \end{tabular}
    \begin{tablenotes}
    \footnotesize
    \item Note: The superscripts $^{***}$, $^{**}$, and $^{*}$ denote the statistical significance at the levels of 1\%, 5\%, and 10\%, respectively.
    \end{tablenotes}
    \end{threeparttable}
  \label{tab:statistics}
\end{table*}

\section{Results and discussion}
\label{S4:EmpAnal}

\subsection{Static quantile spillovers}
\label{S4.1:Static QE}

Table~\ref{Tab:QVAR} reports the static spillover index between food, fossil energy, and clean energy markets. Panel A shows the results estimated at the conditional median ($\tau = 0.5$). Food market is the least affected by other markets, as it predominantly comprises its own spillover effects, accounting for 81.33\% of the total. In addition, food market also makes the least contributions to the other markets, with a proportion of 16.90\%. In contrast, fossil energy market receives the largest impact from the other markets (26.99\%), while the clean energy market is the largest contributor to others (29.43\%). As for the net spillovers, clean energy market stands out as the primary transmitter of spillover effects, which is consistent with the results of \cite{Ahmad-2017-ResIntBusFinanc} and \cite{Saeed-Bouri-Alsulami-2021-EnergyEcon} who reported that return shocks always transmit from clean energy market to fossil energy market. The larger spillover contributions of the clean energy market may reflect its role in the energy transition, which, in turn, influences both fossil energy and food prices. The total return connectedness between these markets is 23.02\%, indicating a moderate level of spillovers between food and energy markets.

\begin{table*}[!t]
  \centering
  \setlength{\abovecaptionskip}{0.1cm}
  \caption{Static return spillovers at different quantiles.}
     \setlength{\tabcolsep}{6.5mm}{
       \begin{threeparttable}
    \begin{tabular}{lcccc}
    \toprule
    & Food Price & Clean Energy & Fossil Energy & From\\
    \midrule
    \multicolumn{4}{l}{\it{Panel A: Median quantile $\tau = 0.5$}}
  \vspace{1mm}
  \\
    Food Price     & 81.33  & 10.76   & 7.91     & 18.67 \\
    Clean Energy   & 8.58   & 76.59   & 14.83    & 23.41 \\
    Fossil Energy  & 8.32   & 18.67   & 73.01    & 26.99\\
    To             & 16.90  & 29.43   & 22.73    & 69.07 \\
    Inc. Own       & 98.24  & 106.02  & 95.74    & \multirow{2}{1cm}{$TSI =$ \\ 23.02} \\
    Net            & -1.76  & 6.02    &-4.26\\
    \midrule
  \multicolumn{4}{l}{\it{Panel B: Extreme lower quantile $\tau = 0.05$}}
  \vspace{1mm}
  \\
    Food Price     & 37.26  & 31.68    & 31.06     & 62.74 \\
    Clean Energy   & 28.39  & 40.00    & 31.61     & 60.00 \\
    Fossil Energy  & 28.62  & 33.07    & 38.31     & 61.69\\
    To             & 57.01  & 64.75    & 62.66     & 184.42 \\
    Inc. Own       & 94.28  & 104.75   & 100.98    & \multirow{2}{1cm}{$TSI =$ \\ 61.47} \\
    Net            & -5.72  & 4.75     & 0.98\\
    \midrule
  \multicolumn{4}{l}{\it{Panel C: Extreme upper quantile $\tau = 0.95$}}\\
  \vspace{1mm}
    Food Price     & 45.57  & 28.30    & 26.13     & 54.43 \\
    Clean Energy   & 28.61  & 41.61    & 29.78     & 58.39 \\
    Fossil Energy  & 29.27  & 31.64    & 39.09     & 60.91 \\
    To             & 57.88  & 59.94    & 55.91     & 173.72 \\
    Inc. Own       & 103.45 & 101.55   & 95.00     & \multirow{2}{1cm}{$TSI =$ \\ 57.91} \\
    Net            & 3.45   & 1.55     & -5.00\\
    \bottomrule
    \end{tabular}
         \begin{tablenotes}
    \footnotesize
    \item Note: `To' indicates the spillover effects that the market transmits to other markets except itself. `Inc. Own' indicates the spillover effects that the market transmits to other markets including itself. `From' indicates the spillover effects of the market received from other markets. `Net' is the disparity between `To' and `From'. `$TSI$' indicates the total spillover index between food, clean energy, and fossil energy markets.
    \end{tablenotes}
    \end{threeparttable}
     }
  \label{Tab:QVAR}
\end{table*}

To explore the spillover effects associate with positive and negative return shocks, we assess the connectedness between these markets at the extreme quantiles ($\tau = 0.05$ and $\tau = 0.95$). The results are presented in Panel B and Panel C of Table~\ref{Tab:QVAR}. Notably, return spillovers at the tails are significantly higher than that at the median. Specifically, the total spillover index is 61.4\% at the extreme lower quantile and 57.91\% at the extreme upper quantile, both of which are considerably higher than the 23.02\% spillover observed at the median. The `To' and `From' indexes at the extreme quantiles are also stronger than those at the median. Moreover, clean energy market remains the net transmitter across all market conditions. Compared to the results at the median, fossil energy market shifts from being a net receiver to a net transmitter at the extremely negative market conditions, while food market changes from a net receiver to a net transmitter under extremely positive shocks.

\begin{figure}[!ht]
\centering
\includegraphics[width=0.6\textwidth,height=0.4\textheight]{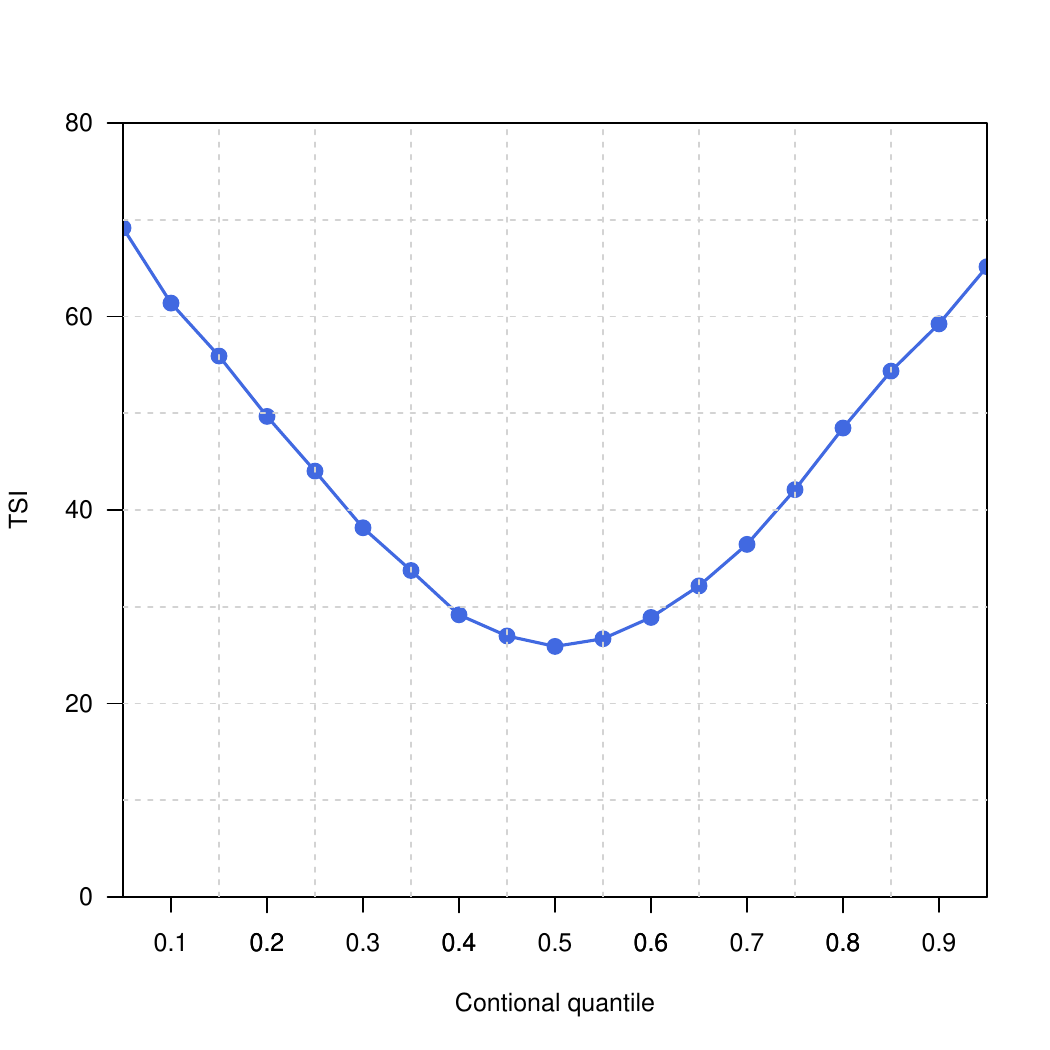}
 \vspace{-10bp}
\caption{Variation in the $TSI$ across various quantiles.} 
\label{Fig:TSI QVAR}
\end{figure}

Moreover, Fig.~\ref{Fig:TSI QVAR} illustrates that the total spillover index at various quantiles follows a U-shape, presenting clear evidence that the total spillover index varies across quantiles and is stronger at the tails. The figure also reveals an asymmetrical pattern, with the index at the left tail being higher than that at the right tail.

\subsection{Dynamic quantile spillovers}
\label{S4.2:Dynamic QE}

To further capture the time-varying characteristics of the connectedness between food and energy markets, we estimate the dynamic spillover effects using the rolling window method, with window size of 36 and forecast horizon of 12.

\begin{figure}[!ht]
\centering
\includegraphics[width=0.475\textwidth]{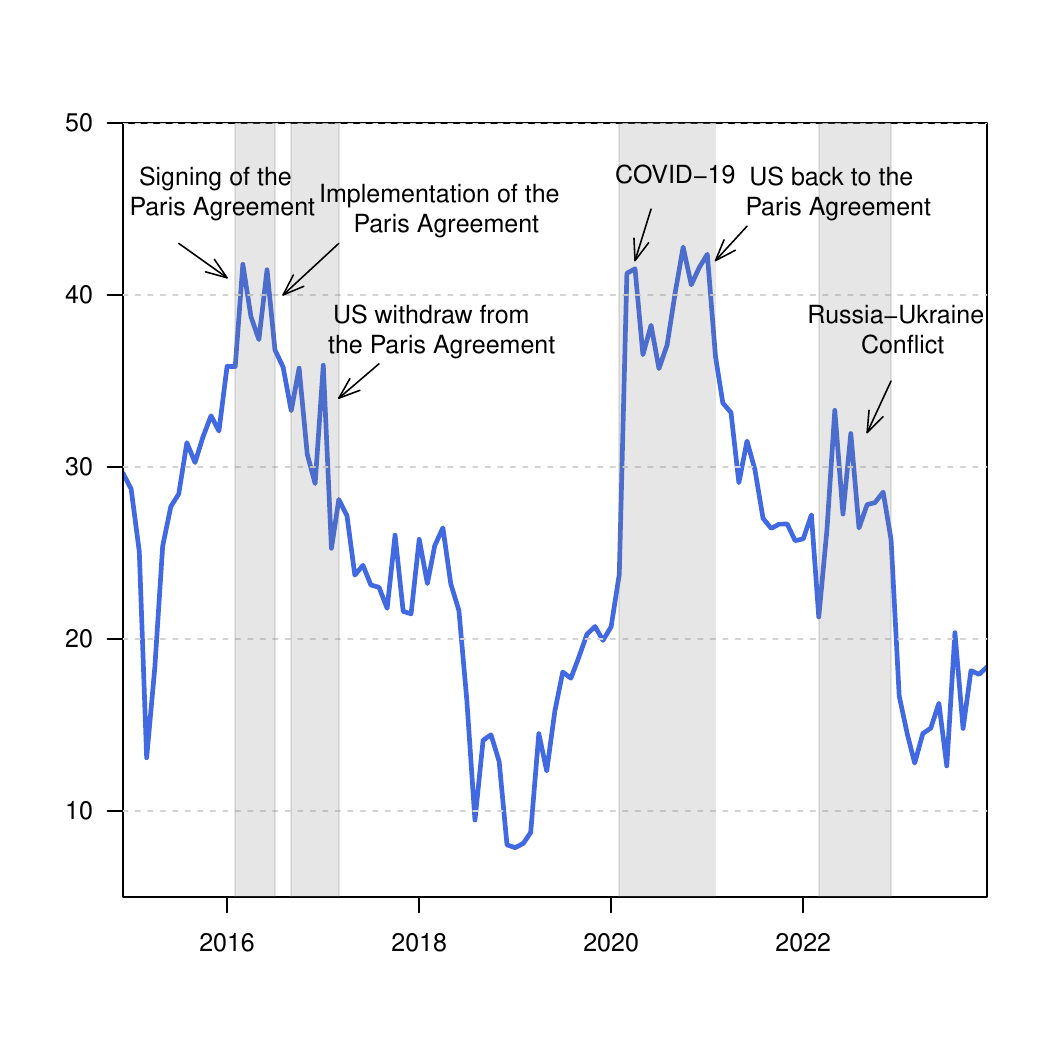}
\includegraphics[width=0.475\textwidth]{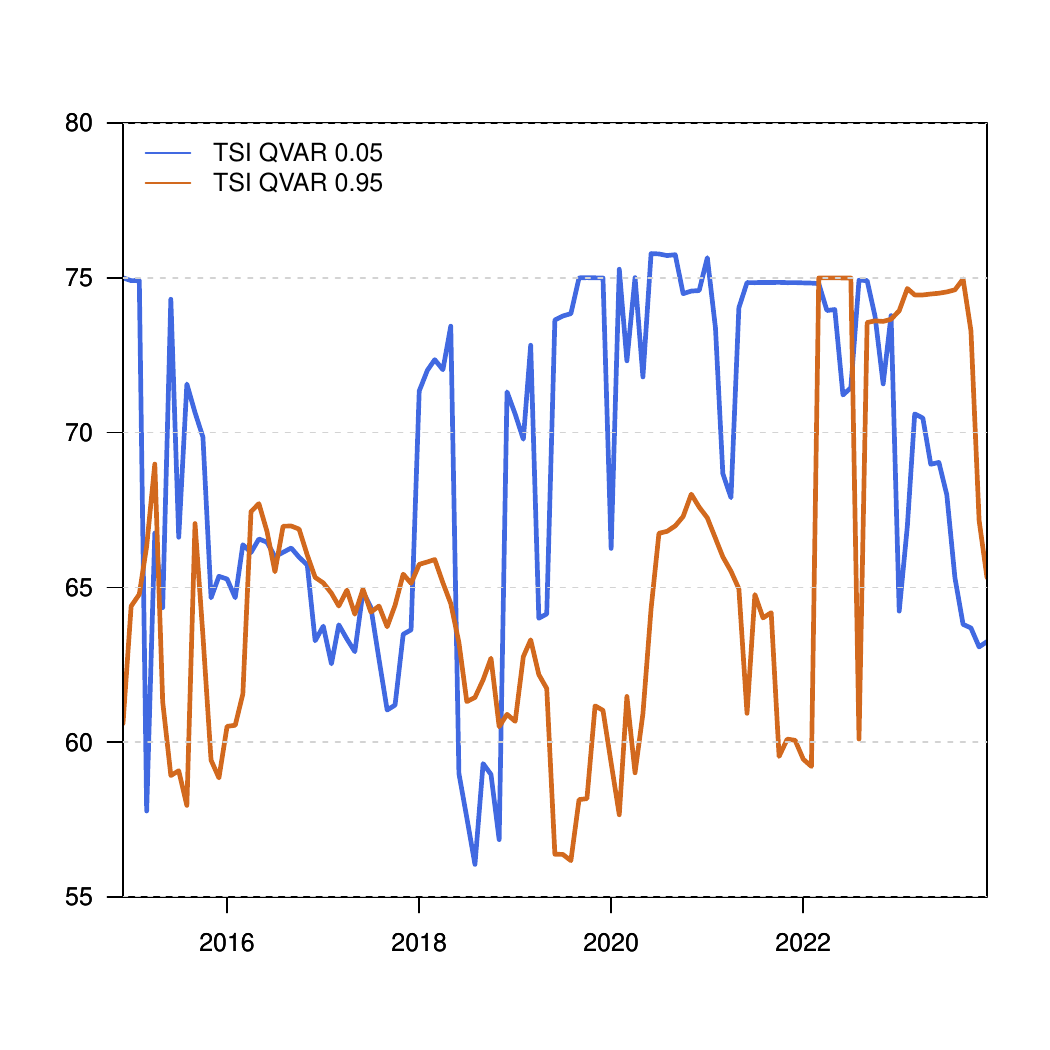}
 \vspace{-25bp}
\caption{Total return spillovers for median quantile $\tau = 0.5$ (left panel) and extreme lower and upper quantiles $\tau = 0.05$ and $\tau = 0.95$ (right panel).} 
\label{Fig:TSI:3quantiles}
\end{figure}

The left panel of Fig.~\ref{Fig:TSI:3quantiles} shows that the total spillovers estimated at the median quantile, which fluctuate between 7.87\% and 42.79\%, with a standard deviation of 8.95. Moreover, the variation trend of the spillover index indicates that the connectedness between food and energy markets increases significantly during extreme events, such as the signing and implementation of the Paris Agreement in 2015 and 2016, the withdrawal of the US from the Paris Agreement in 2017 and its return in early 2021, the COVID-19 pandemic in 2020, and the Russia-Ukraine conflict in 2022. This result is consistent with the finding of \cite{Cao-Xie-2024-IntJFinancEcon} that extreme events strengthen the connectedness between markets. Furthermore, we analyze the dynamic spillovers between these markets at the extreme quantiles, and the results are presented in the right panel of Fig.~\ref{Fig:TSI:3quantiles}. The total spillovers at the tails are substantially higher compared to the median. The total spillover index fluctuates less at the tails, varying between 56.18\% and 75.00\% with standard deviation of 5.15 at the right tail and between 56.05\% and 75.78\% with standard deviation of 5.39 at the left tail. These findings highlight that spillovers are not only stronger at the extremes but also more stable in comparison to the median, underscoring the heightened market interdependence during periods of significant positive or negative shocks.

\begin{figure}[!h]
\centering
\includegraphics[width=0.8\textwidth,height=0.4\textheight]{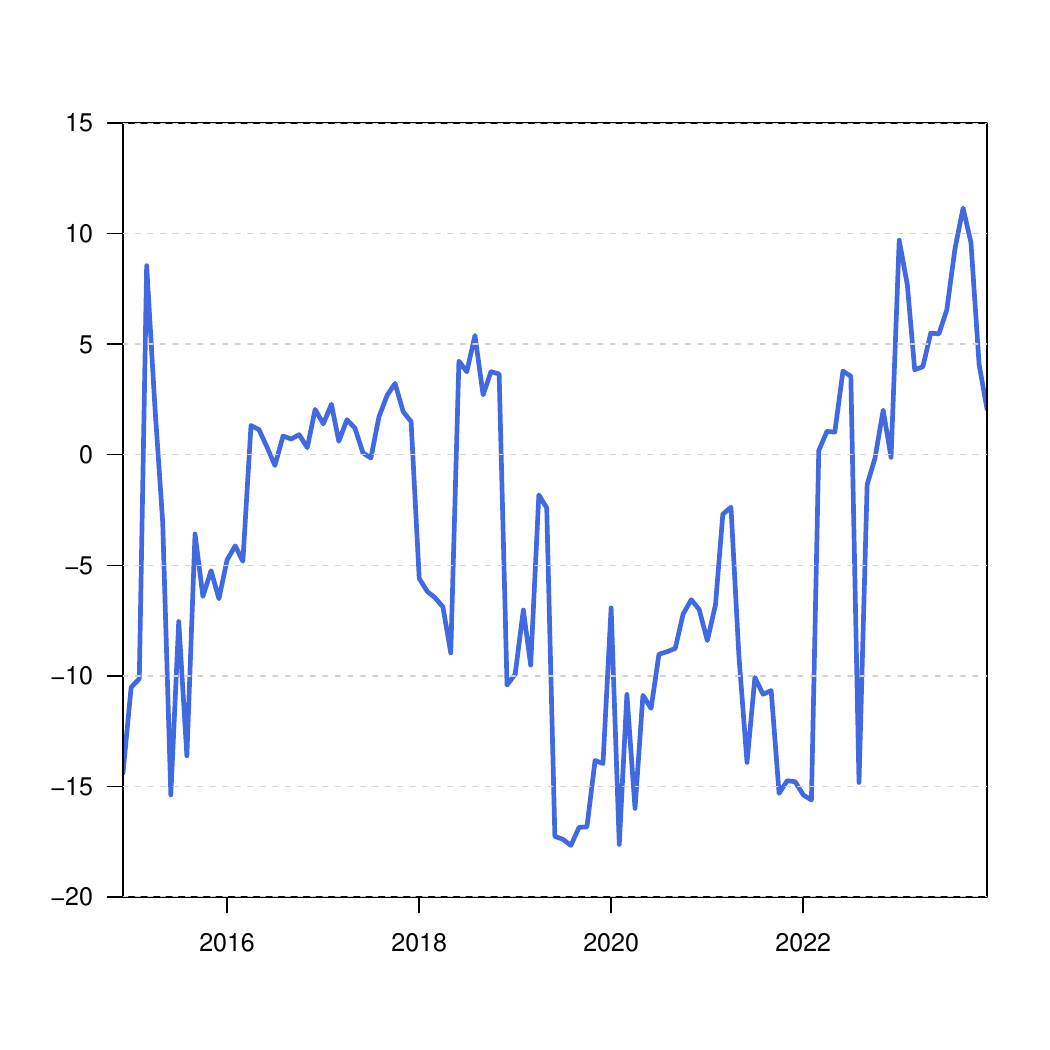}
 \vspace{-25bp}
\caption{Relative tail dependence ($TSI_{\tau = 0.95}-TSI_{\tau = 0.05}$).} 
\label{Fig:RTD}
\end{figure}

To assess the potential presence of asymmetry as shown in Fig.~\ref{Fig:TSI QVAR}, we calculate the relative tail dependence ($RTD$, $TSI_{\tau=0.95}-TSI_{\tau=0.05}$) \citep{Saeed-Bouri-Alsulami-2021-EnergyEcon}. Fig.~\ref{Fig:RTD} shows that 5\% $RTD$ varies between positive and negative values. A greater proportion of the values are negative, indicating that the spillovers are stronger at the left tail than at the right tail.

\begin{figure}[!h]
   \centering
	\includegraphics[width=0.321\linewidth]{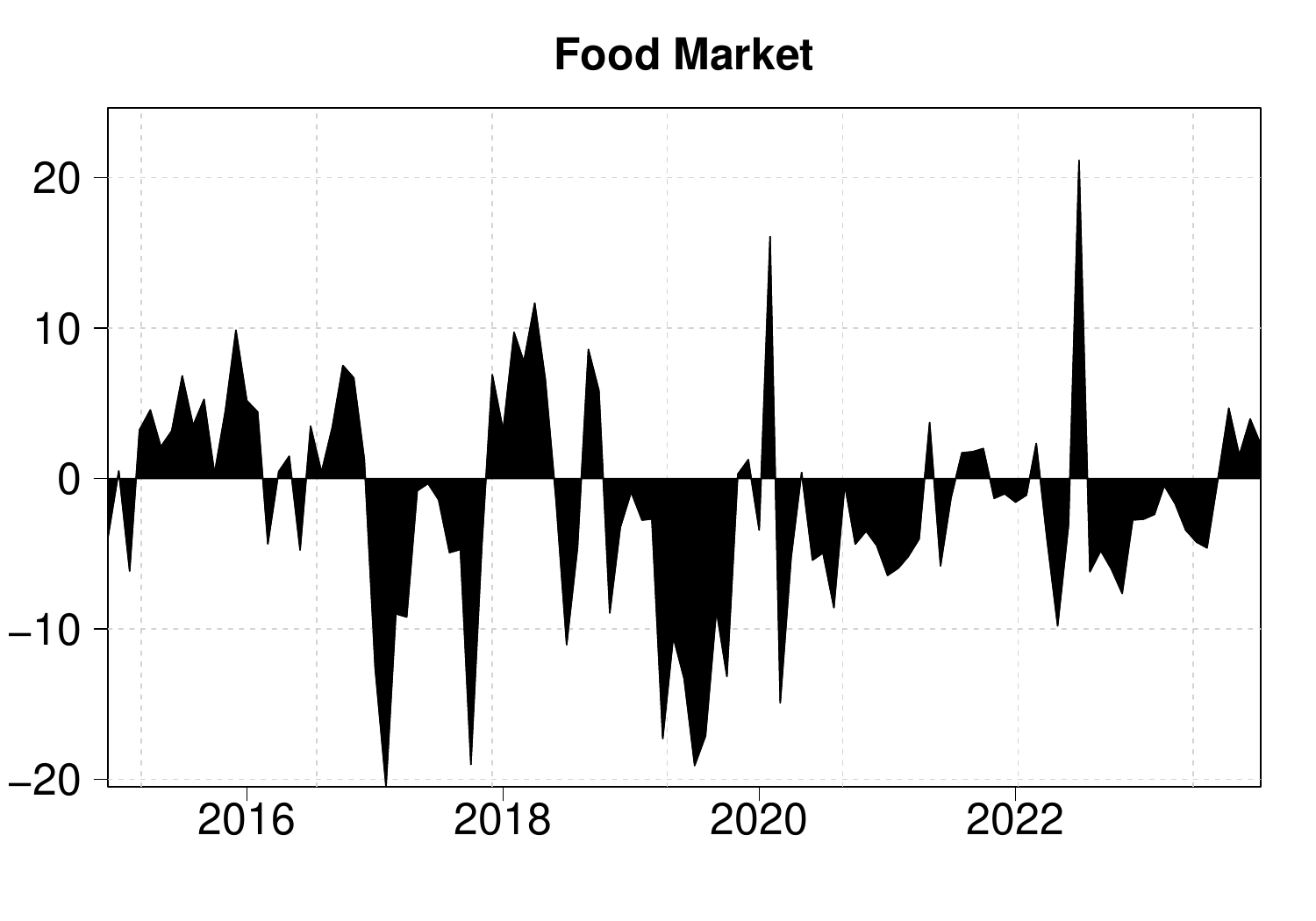}
	\includegraphics[width=0.321\linewidth]{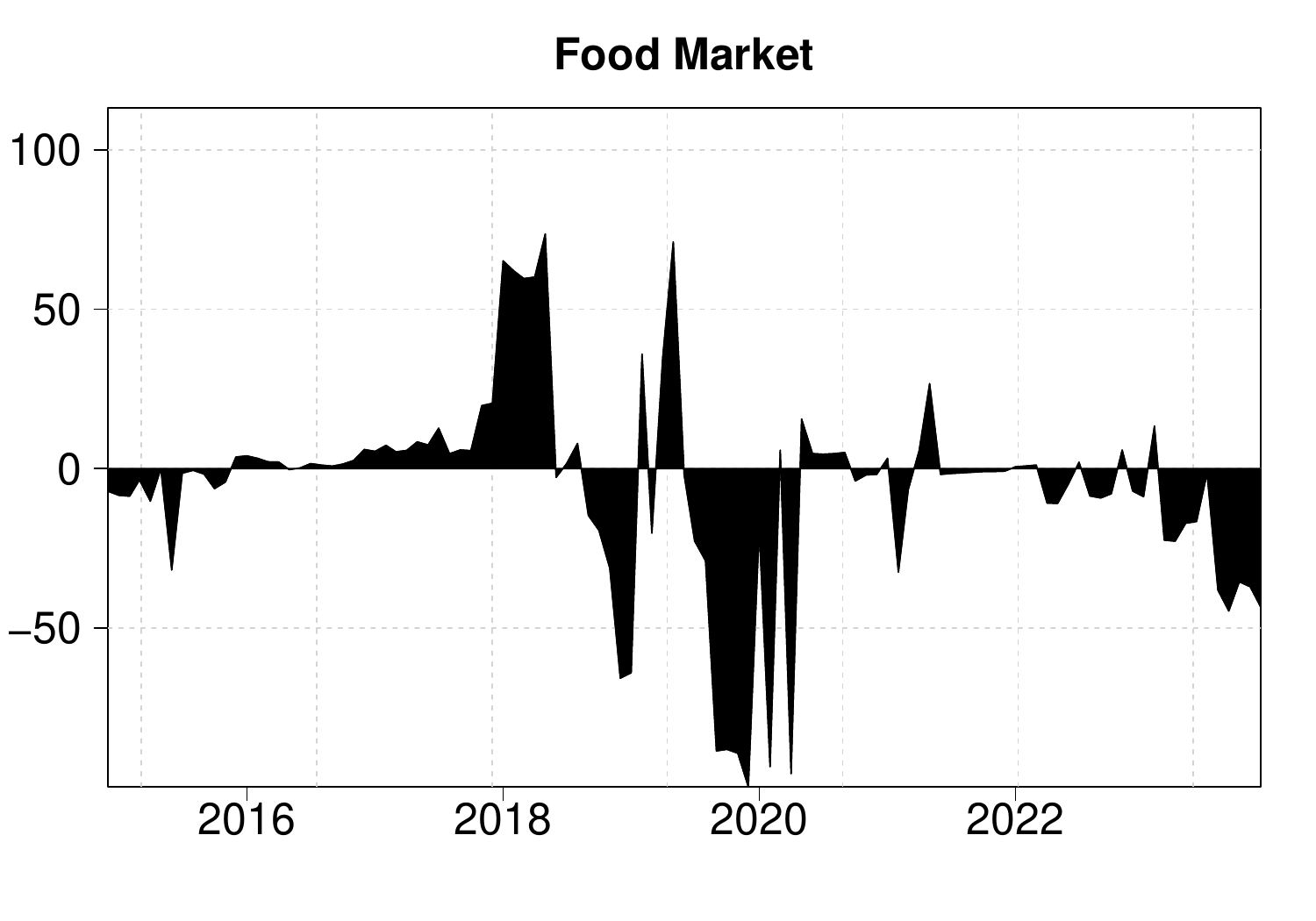}
	\includegraphics[width=0.321\linewidth]{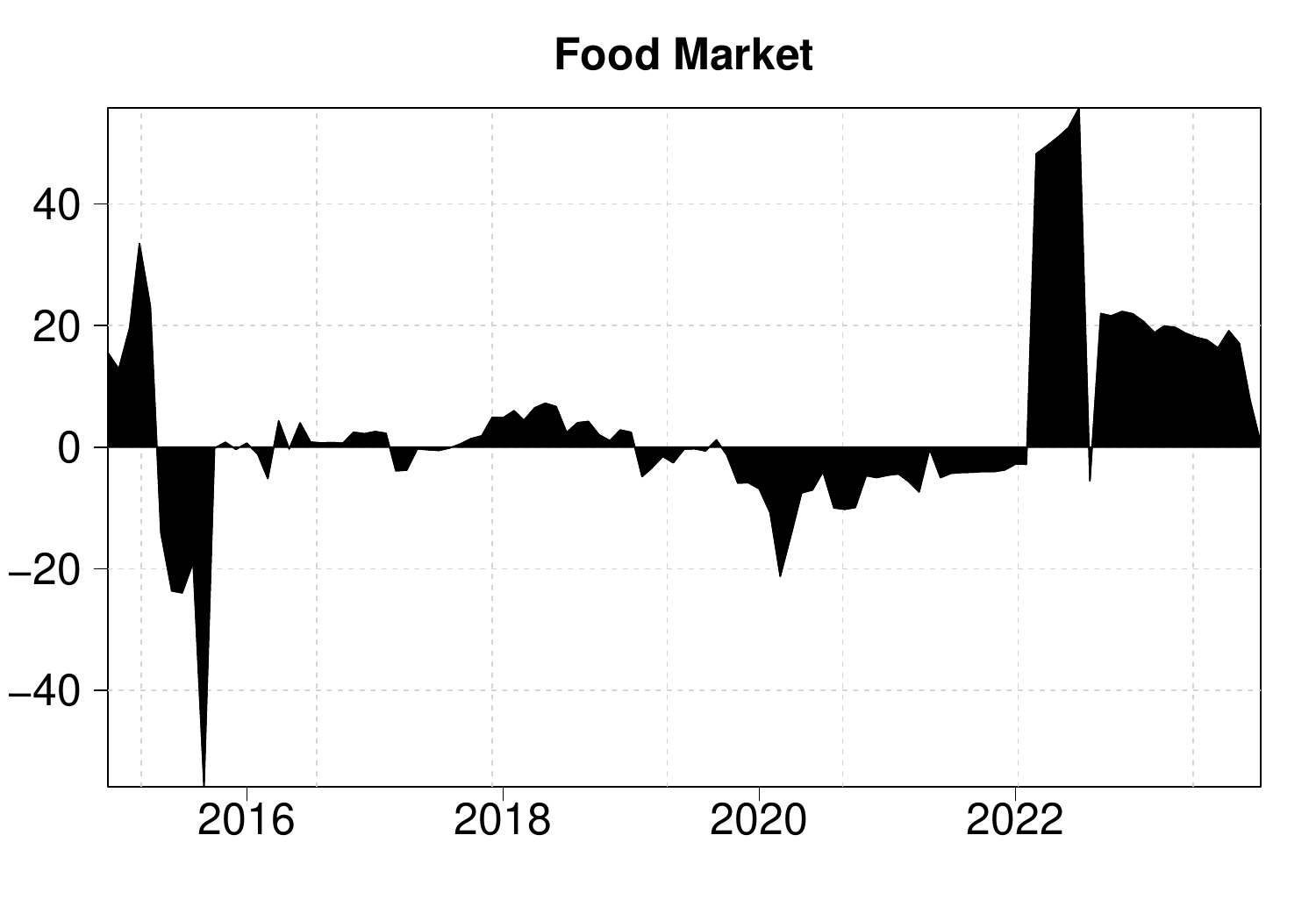}
	\includegraphics[width=0.321\linewidth]{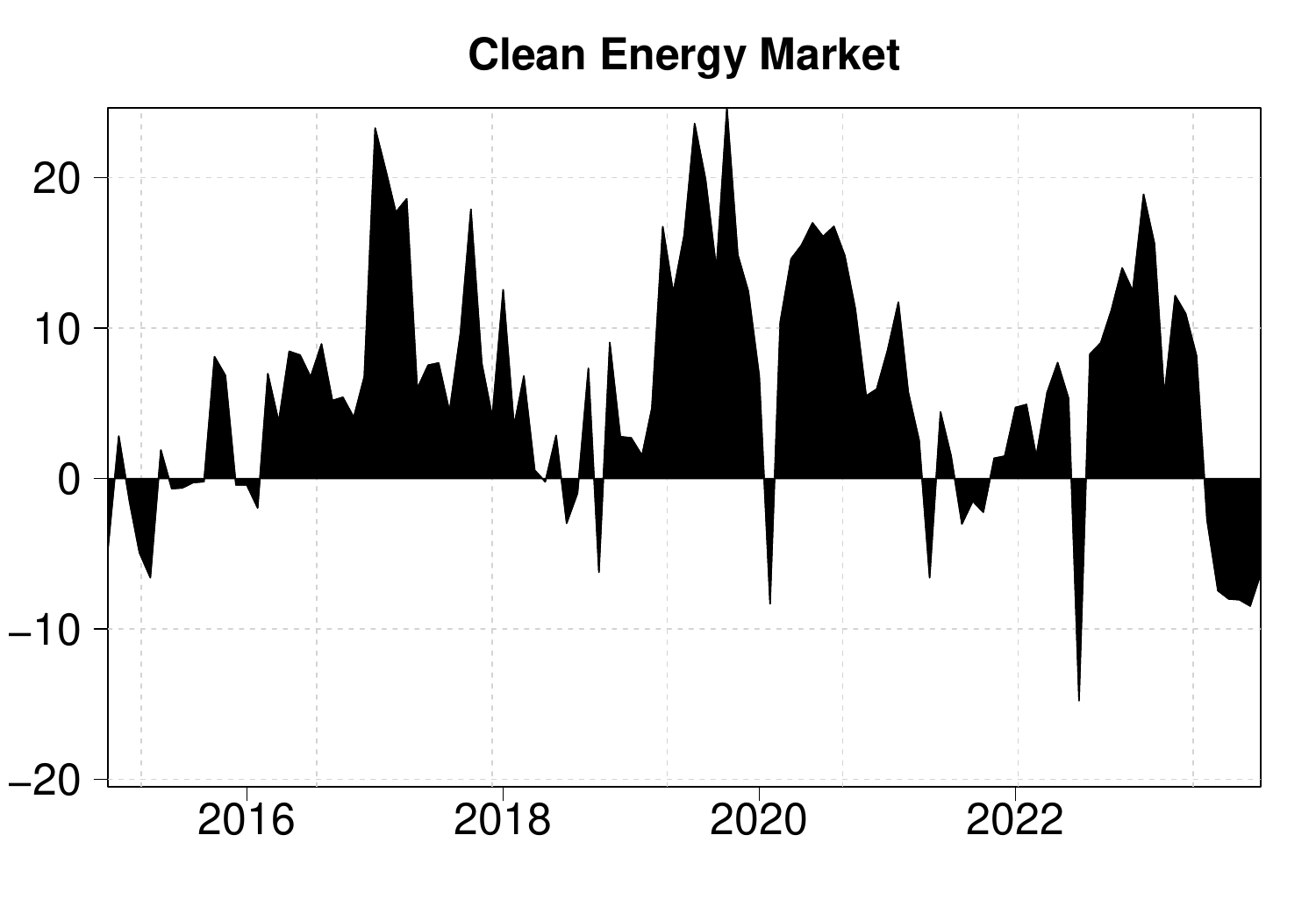}
	\includegraphics[width=0.321\linewidth]{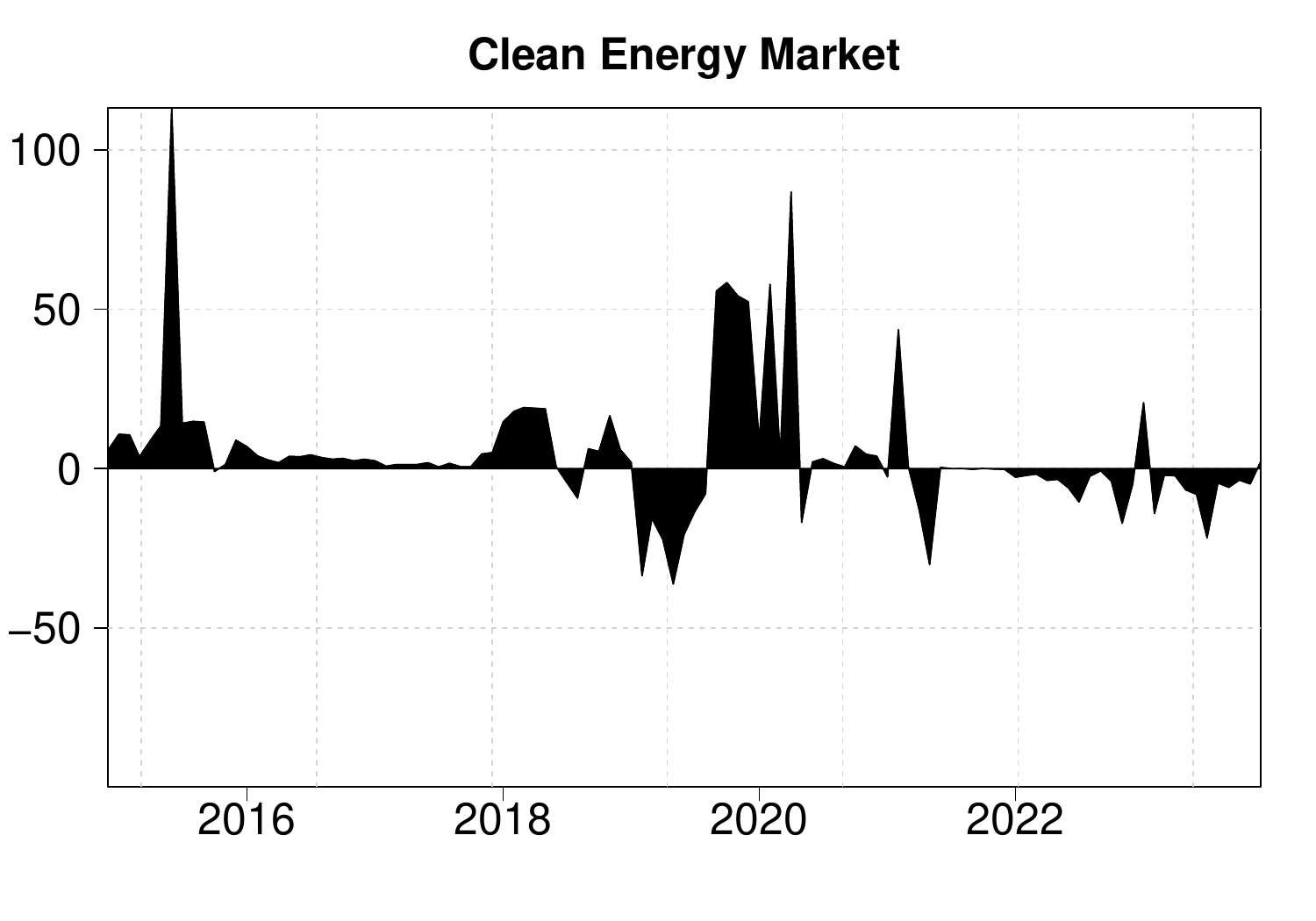}
	\includegraphics[width=0.321\linewidth]{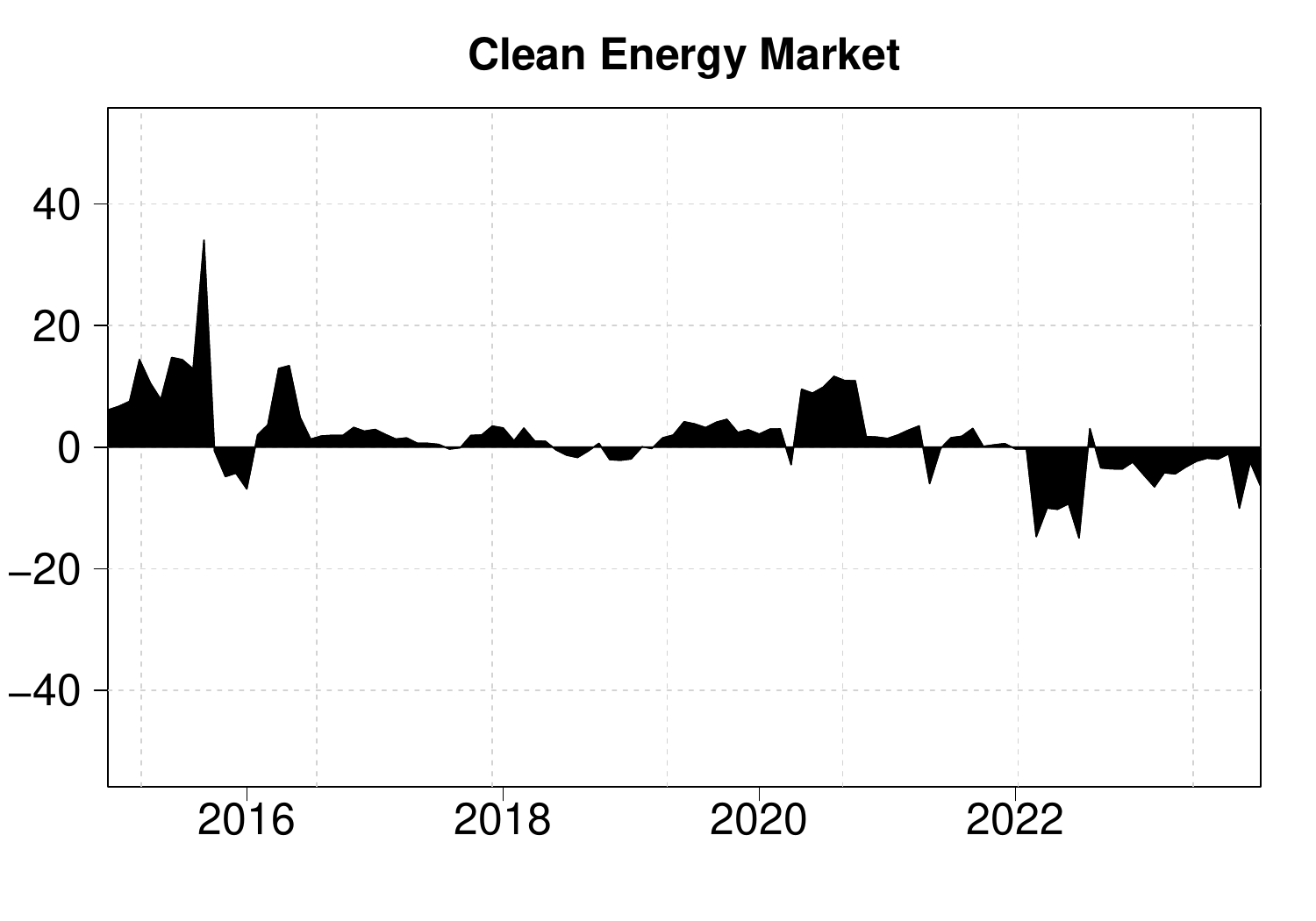}
	\includegraphics[width=0.321\linewidth]{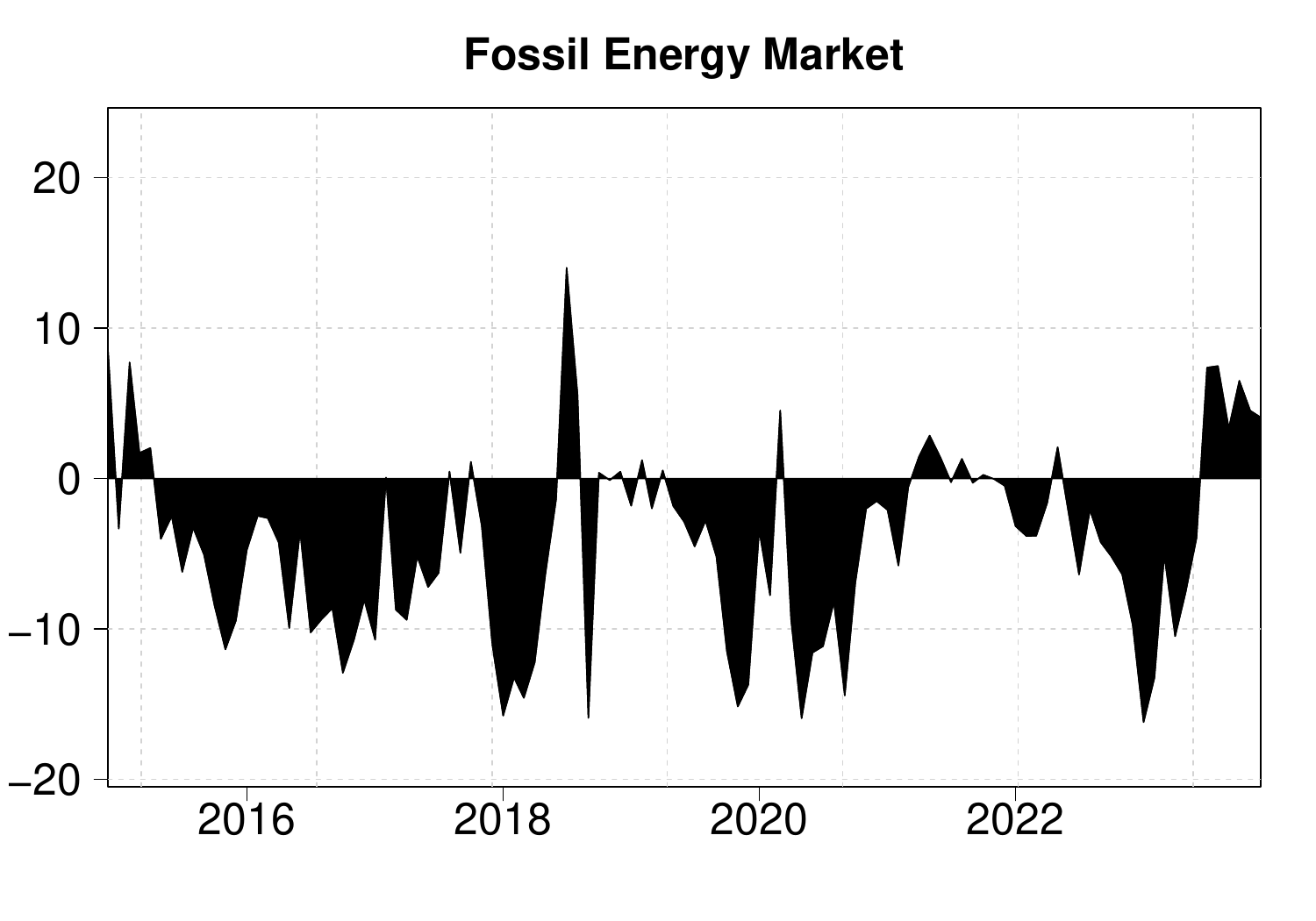}
	\includegraphics[width=0.321\linewidth]{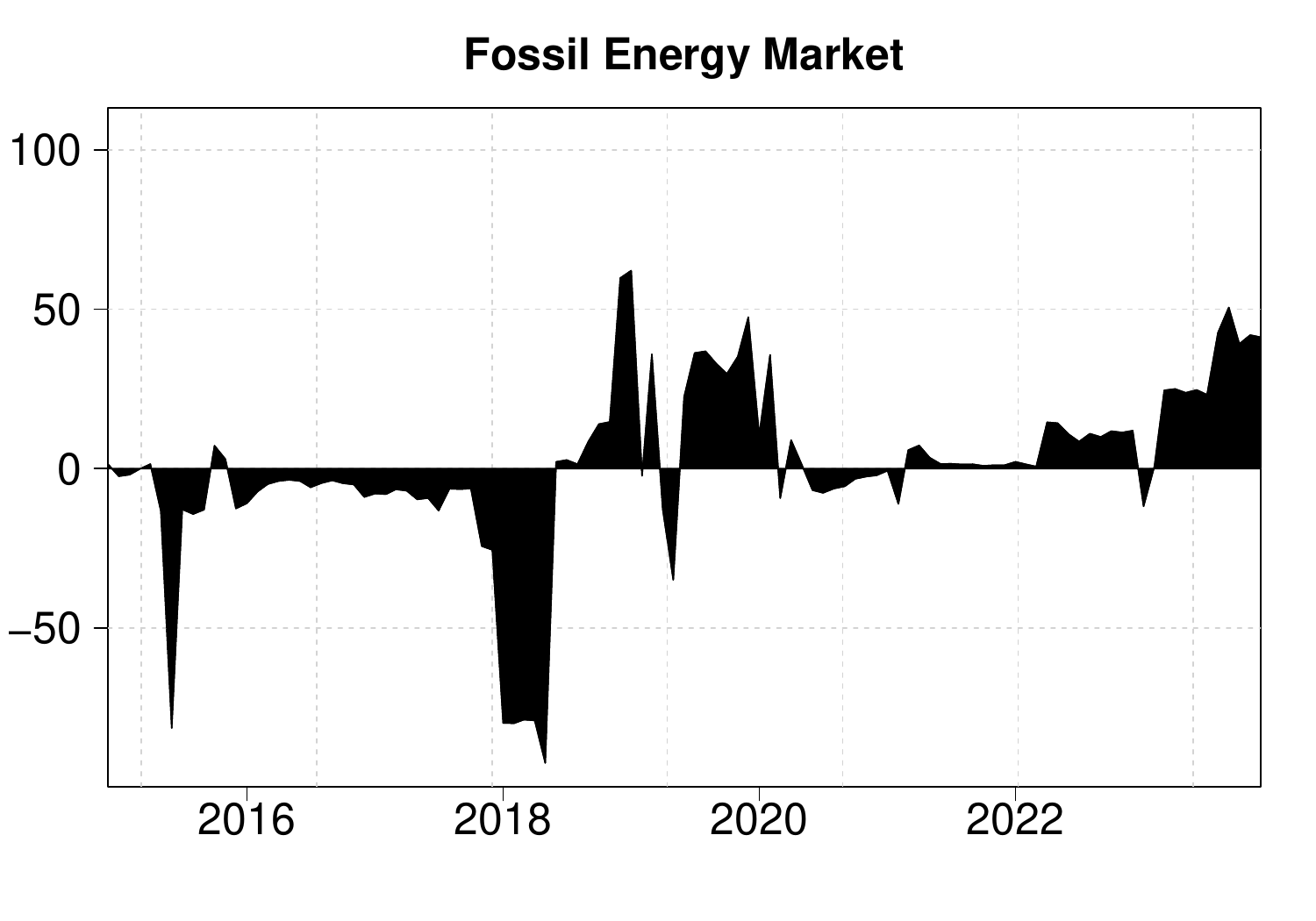}
	\includegraphics[width=0.321\linewidth]{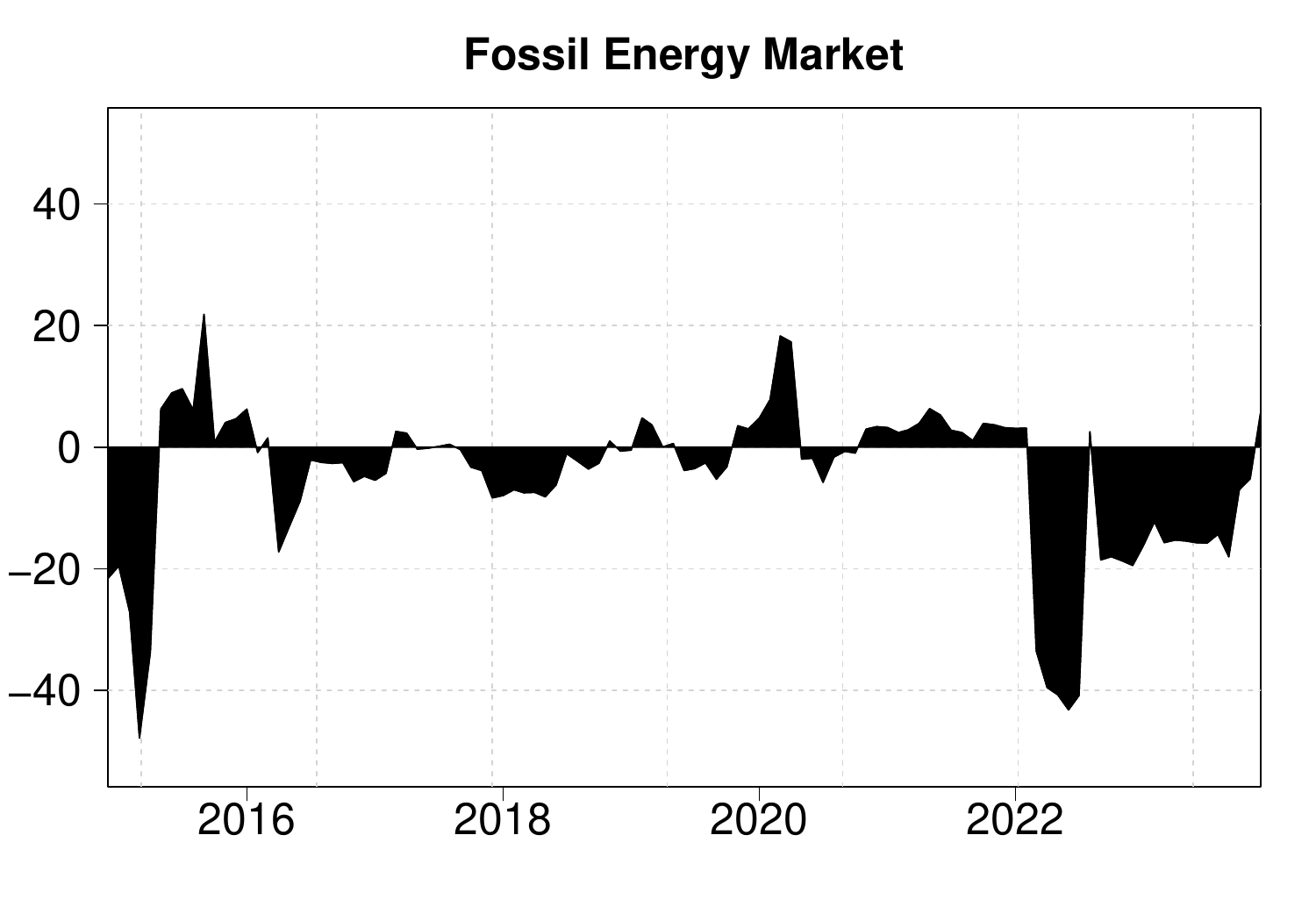}
    \caption{Net return spillovers. The left, meddle, and right panels correspond to $\tau = 0.5$, $\tau = 0.05$, and $\tau = 0.95$, respectively.} 
    \label{Fig:NSI}
\end{figure}

The net spillover effects estimated at the median, extreme upper, and extreme lower quantiles are shown in the left, middle, and right panels of Fig.~\ref{Fig:NSI}, respectively. The left panel shows that food market alternates between serving as a net transmitter and a net receiver. Clean energy market acts as a net transmitter, as its net spillover index is positive throughout most of the sample period. Conversely, the net return spillover index for fossil energy market are predominantly negative, indicating they are net recipients. This aligns with the view that the spillover effects from the clean energy market to the other markets increase as energy consumption shifting from fossil fuels to clean energy \citep{Raza-Khan-Benkraiem-Guesmi-2024-IntRevFinancAnal}. The median and right panels of Fig.~\ref{Fig:NSI} show that the patterns of net spillovers are not identical for the left and right tails. The estimates fluctuate significantly at tails, indicating that fossil energy, clean energy, and food markets are changing between net transmitters and net recipients under extreme market conditions.

\subsection{Robustness tests}
\label{S4.3:Robust}
\begin{figure}[!h]
\centering
\includegraphics[width=0.49\textwidth,height=0.33\textheight]{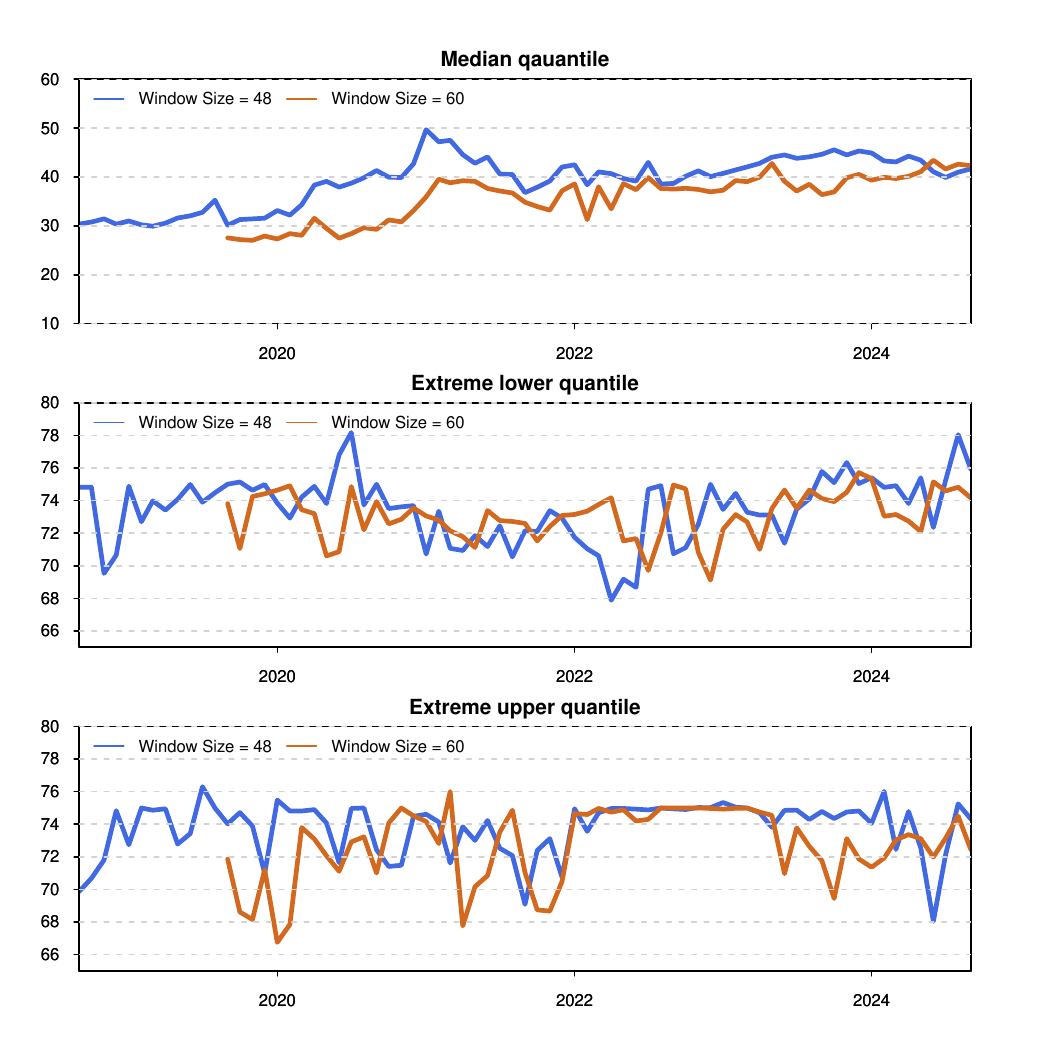}
\includegraphics[width=0.49\textwidth,height=0.33\textheight]{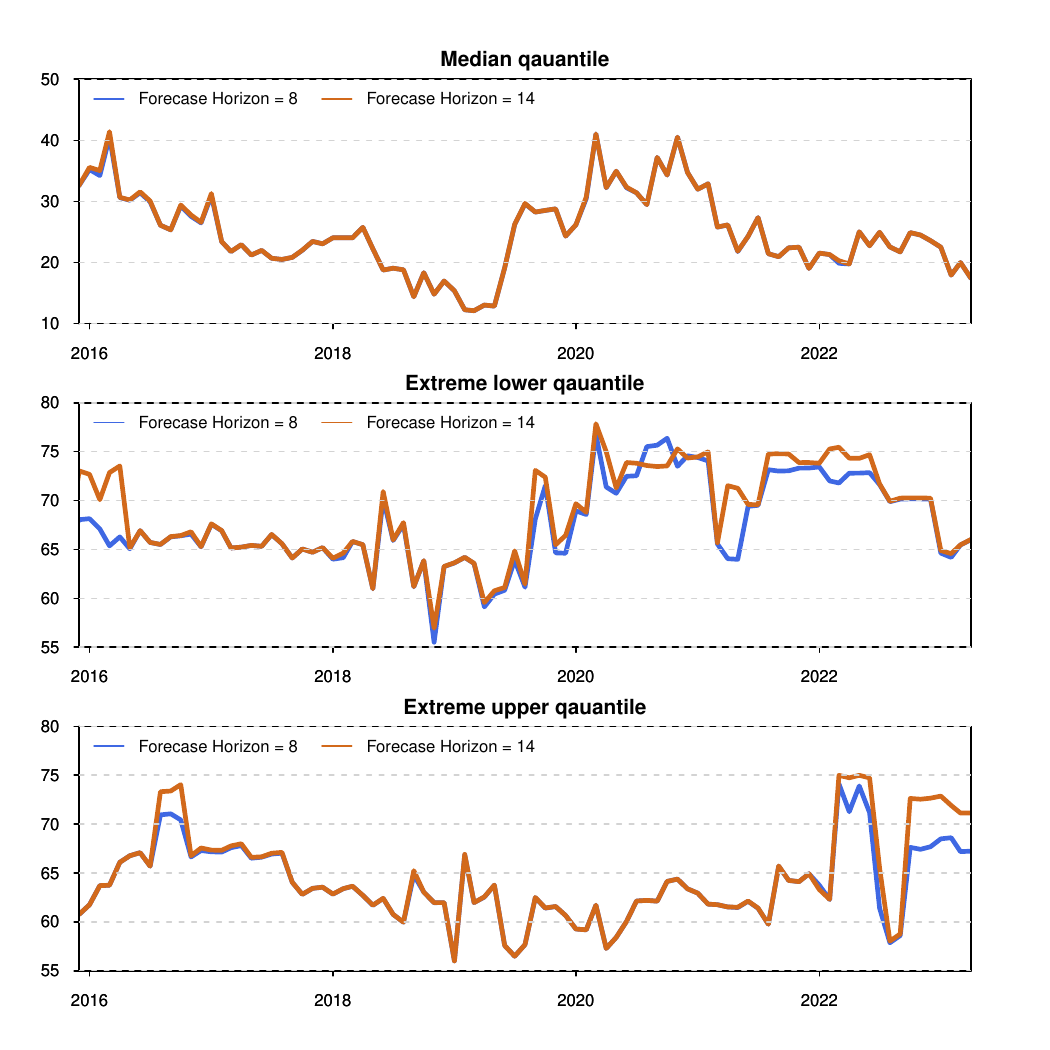}
\caption{Total return spillovers in quantile VAR. Left: Window size = 48 or 60, forecast horizon = 12; Right: Window size = 36, forecast horizon = 8 or 14.} 
\label{Fig:WindowSize:Horizon}
\end{figure}

On the one hand, we assess the robustness of the aforementioned results by varying the rolling-window size and the forecast horizon. First, we consider window sizes of 48 months or 60 months while keeping the forecast horizon fixed at 12. The results, as reported in left panel of Fig.~\ref{Fig:WindowSize:Horizon}, show that the pattern of the spillover is not shaped by the window size. Second, we adjust the forecast horizon to 8 or 14. The results presented in the right panel of Fig.~\ref{Fig:WindowSize:Horizon} show that the spillovers are still robust when the forecast horizon is changed. 

\begin{figure}[!h]
\centering
\includegraphics[width=0.8\textwidth,height=0.4\textheight]{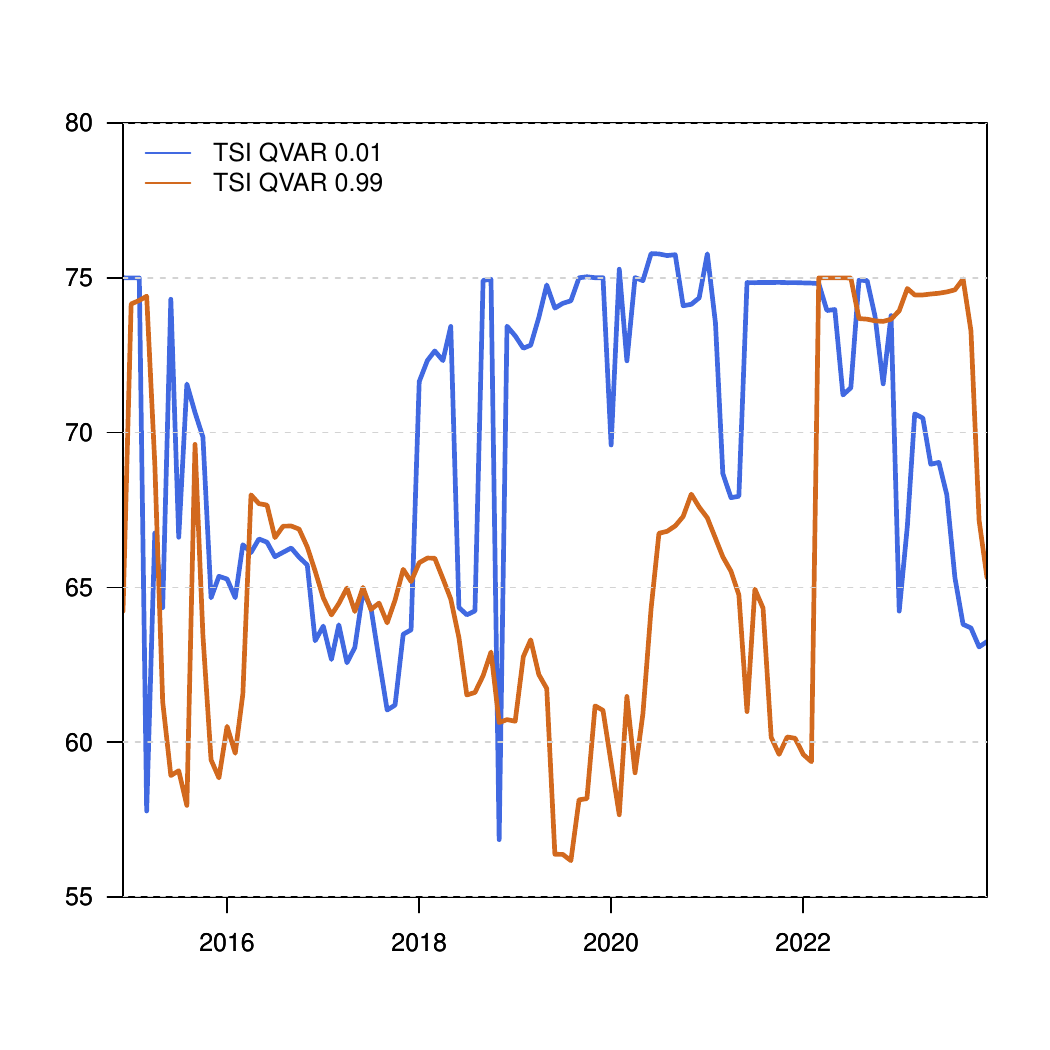}
\caption{Total return spillovers in quantile VAR (Extreme lower quantile $\tau = 0.01$ and extreme upper quantile $\tau = 0.99$.)} 
\label{Fig:001099quantile}
\end{figure}

On the other hand, we examine the 0.01 and 0.99 quantiles for extreme positive and negative conditions, respectively. When compared with the right panel of  Fig.~\ref{Fig:TSI:3quantiles}, the results in Fig.~\ref{Fig:001099quantile} for the 1\% extreme quantiles are similar to the previous trends.

\section{The role of external uncertainties}
\label{S5:External:Uncertainties}

According to Fig.~\ref{Fig:TSI:3quantiles}, the total spillover indexes are prominently affected by the external uncertainties, such as climate policy uncertainty (e.g., the signing and implementation of the Paris Agreement) and geopolitical risks (e.g., the Russia-Ukraine Conflict). In this section, we examine the impact of external uncertainties on the spillover effects between food and energy markets.

We examine five key types of external uncertainties in this study\footnote{We obtain the data from \url{https://www.policyuncertainty.com}.}. The first is economic policy uncertainty (EPU), which is computed from the GDP-weighted average of 21 nations' economic policy uncertainty indices \citep{Baker-Bloom-Davis-2016-QJEcon}. Researchers conclude that EPU complicates the food-energy cross-market return spillovers through direct and indirect channels \citep{Cao-Cheng-Li-2023-ResourPolicy}. On the one hand, from the commodity attributes of food and energy, high EPU leads producers to lower investments and also reduces demands for commodities as raw materials. On the other hand, from the financial properties, high EPU prompts investors to hedge risks by investing financialized commodity markets. The second type is climate policy uncertainty (CPU), which is developed by analyzing newspaper articles on climate \citep{Engle-Giglio-Kelly-Lee-Stroebel-2020-RevFinancStud}.The nexus between CPU and energy markets stems from the fact that climate policies have boosted the clean energy industry since the government introduced development goals and policies aimed at reducing greenhouse gas emissions, stimulating investors to switch the investment from traditional fossil energy to clean energy market \citep{Uddin-Sahamkhadam-Yahya-Tang-2023-IntJFinancEcon,Syed-Apergis-Goh-2023-Energy}. For the CPU-food market nexus, climate policy works on the global climate environment and then affects food production \citep{Liu-Luo-Xu-Zhang-2023-FinancResLett,Chandio-Jiang-Akram-Ozturk-Rauf-Mirani-Zhang-2023-IntJFinancEcon}. The third one is trade policy uncertainty, which is constructed by integrating information related to trade policy on US newspaper articles \citep{Baker-Bloom-Davis-2016-QJEcon}. In the context of global trade war, the connectedness between food and energy market are affected by channels of commodity trading and risk hedging \citep{Mei-Xie-2022-FinancResLett,Yang-Zhang-Qin-Niu-2024-ResIntBusFinanc}. Geopolitical risk (GPR) is the forth uncertainty since food, fossil energy, and clean energy markets are sensitive to both geopolitical threats and acts \citep{Yousfi-Bouzgarrou-2024-EnvironSciPollutRes}. Lastly, we introduce COVID-19 as a dummy variable, where it takes a value of 1 between January 2020 and December 2020 and 0 otherwise.


We reveal the effect of these external uncertainties on the total spillovers at the conditional median ($TSI_{\tau=0.5}$), extreme lower quantile ($TSI_{\tau=0.05}$), and extreme upper quantile ($TSI_{\tau=0.95}$). First of all, we examine the existence of unit roots for each variable by using ADF test \citep{Dickey-Fuller-1979-JASA}, PP test \citep{Phillips-Perron-1988-Bm}, and KPSS test \citep{Kwiatkowski-Phillips-Schmidt-Shin-1992-JEm}. As the results presented in Table~\ref{Tab:UnitRootTest}, all variables are either $I(0)$ or $I(1)$ processes. It is find that the application of ARDL and NARDL models is valid. 

\subsection{Results of ARDL models}
\label{S5.1:ARDL}

According to the Akaike information criterion, we obtain ($n_1, n_2, n_3, n_4, n_5$) as (1, 1, 4, 1, 2, 0), (1, 1, 1, 3, 3, 0), and (1, 3, 3, 1, 2, 0) in ARDL models when the dependent variable is $TSI_{\tau=0.5}$, $TSI_{\tau=0.05}$, and $TSI_{\tau=0.95}$, respectively. Results of ARDL models are presented in Table~\ref{Tab:ARDL}. The statistics of bound $F$ test shown as $F_{PSS}$ are 3.382, 4.265, and 4.252 for $TSI_{\tau=0.5}$, $TSI_{\tau=0.05}$, and $TSI_{\tau=0.95}$, respectively. They are all significant at either 10\% or 5\% levels, indicating the existence of long-run cointegration between the spillovers and external uncertainties under both normal and extreme market conditions. 

For $TSI_{\tau=0.5}$, the short-run results show that contemporaneous $\Delta ln EPU$ has a positive effect on $\Delta ln TSI_{\tau=0.5}$, while $\Delta ln CUP$, $\Delta ln TUP$, and $\Delta ln GPR$ each has a negative effect. However, the contemporaneous effects are not statistically significant, as indicated by the corresponding $P$-values. The coefficients for the first and third lags of $\Delta ln CUP$ are significantly positive. The magnitude of coefficients indicate that a 1\% increase in  the first and third lags of $ln CPU$, leads to a 0.096\% and 0.112\% increase in $ln TSI_{\tau=0.5}$, respectively. The coefficient of first lag of $ln GPR$ is significantly negative, with a magnitude of 0.176\%. The long-run results are reported in Panel B of Table~\ref{Tab:ARDL}. $ln CPU$ has a significantly negative long-run impact on $ln TSI_{\tau=0.5}$, with a magnitude of 0.548\%, while $ln GPR$ and $COVID-19$ have significantly positive long-rung impacts, with magnitudes of 0.712\% and 0.969\%, respectively. 

We also examine the relationships between spillovers and external uncertainties at the extreme quantiles. For $TSI_{\tau=0.05}$, the coefficients for $\Delta ln EPU$, $\Delta ln CPU$, $\Delta ln TPU$, and $\Delta ln GPR$ are all positive, but only significant for $\Delta ln EPU$ and $\Delta ln GPR$. specifically, a 1\% increase in $ln EPU$ and $ln GPR$ contributes to the change of $ln TSI_{\tau=0.05}$ in 0.048\% and -0.051\%, respectively. The coefficients of first and second lags of $\Delta ln TPU$ and $\Delta ln GPR$ are significantly positive. A 1\% increase in the first and second lags of $ln TPU$ results in change of 0.017\% and 0.015\% in $ln TSI_{\tau=0.05}$, while the percentages for the first and second lags of $ln GPR$ are 0.046\% and 0.051\%. The long-run results reveal that $ln EPU$ has a significantly positive impacts to the total spillover at the lower quantile with a magnitude of 0.137, while $ln GPR$ has a significantly negative impact with a magnitude of 0.205. For the spillover at the upper quantile, contemporaneous $\Delta ln EPU$ and $\Delta ln TPU$ have coefficients of 0.074 and -0.02 respectively, with statistical significance at 1\% level. Moreover, the first and second lagged $\Delta ln EPU$ and the first lagged $\Delta ln GPR$ have significantly negative impacts on $ln TSI_{\tau=0.95}$, while second lagged $\Delta ln CPU$ has a significant positive impact. It is noted that all of the coefficients of $ECT$ are significantly negative. The magnitudes of the coefficients of $ECT$ imply the speed of adjustment toward long-run equilibrium from the short-run uncertainty shocks.

Additionally, We take residual diagnostics tests to approve the adequacy of the selected ARDL models, including the Breusch-Godfrey LM test \citep{Breusch-1978-AustEconPap, Godfrey-1978-Econometrica} for the autocorrelation, the Breusch-Pagan test \citep{Breusch-Pagan-1979-Econometrica} for the heteroskedasticity, the Ramsey RESET statistics test regress specification error \citep{Ramsey-1969-JRStatSocB} for the normality, and CUSUM 
test \citep{Brown-Durbin-Evans-1975-JRStatSocB} for the model stability. The results outlined in Panel C of Table~\ref{Tab:ARDL} provide the evidence that models pass these diagnostics except for the heteroskedasticity and normality of models for $TSI_{\tau=0.5}$.

\begin{table*}[!t]
  \centering
  \setlength{\abovecaptionskip}{0.1cm}
  \caption{Results of conventional unit root tests}
     \setlength{\tabcolsep}{0.2mm}{
       \begin{threeparttable}
    \begin{tabular}{lcccccc}
    \toprule
      \multirow{2}{1.5cm}{Variable}& \multicolumn{2}{c}{ADF} & \multicolumn{2}{c}{PP} & \multicolumn{2}{c}{KPSS}\\
      \cmidrule(r){2-3} \cmidrule(r){4-5}\cmidrule(r){6-7}
       & Intercept & Trend \& intercept & Intercept & Trend \& intercept & Intercept & Trend \& intercept \\
    \midrule
    \multicolumn{7}{l}{\textit{Panel A: Level}}\\
    $lnTSI_{\tau=0.5}$  & -1.1794$^{~~~}$    & -3.4292$^{*~~}$    & -1.1794$^{~~~}$   &-3.5138$^{**~}$   & 1.1230$^{***}$
 &  0.1801$^{**~}$\\
    $lnTSI_{\tau=0.05}$  & -4.1716$^{***}$    & -4.5447$^{***}$   & -4.2001$^{***}$ &-4.6347$^{***}$   & 0.4845$^{**~}$ &  0.1599$^{**~}$ \\
    $lnTSI_{\tau=0.95}$  & -3.3966$^{**~}$  & -3.8563$^{**~}$ & -3.2652$^{**~}$   & -3.7664$^{**~}$   & 0.4882$^{**~}$ & 0.1942$^{**~}$\\
    $lnEPU$  & -2.2226$^{~~~}$    &-4.7694$^{***}$  & -2.5217$^{~~~}$  & -4.6378$^{***}$   & 1.1591$^{***}$ & 0.1172$^{~~~}$\\
    $lnCPU$  & -10.0305$^{***}$ $^{~}$ & -10.2271$^{***}$  & -10.0952$^{***}$   & -10.2163$^{***}$ 
   & 0.3140$^{~~~}$ & 0.0595$^{~~~}$ \\
    $lnTPU$  & -2.5755$^{~~~}$    & -2.4557$^{~~~}$    & -3.9257$^{***}$  & -3.9042$^{**~}$  & 0.3782$^{*~~}$
 & 0.2721$^{***}$ \\
    $lnGPR$  & -5.1127$^{***}$    & -5.4250$^{***}$   & -5.0112$^{***}$  & -5.3600$^{***}$   & 0.3014$^{~~~}$ & 0.1211$^{*~~}$\\ 
    \midrule
    \multicolumn{7}{l}{\textit{Panel B: First difference}}\\
    $\Delta lnTSI_{\tau=0.5}$  & -11.6171$^{***}$  & -11.6192$^{***}$ & -11.6145$^{***}$   & -11.6167$^{***}$   & 0.1028 & 0.0794\\
    $\Delta lnTSI_{\tau=0.05}$  & -14.0546$^{***}$  & -13.9868$^{***}$ & -17.6571$^{***}$   & -17.5518$^{***}$   & 0.1188 & 0.1187\\
    $\Delta lnTSI_{\tau=0.95}$  & -7.2782$^{***}$  & -7.1848$^{***}$ & -14.1296$^{***}$   & -14.0499$^{***}$   & 0.0576 & 0.0581\\
    $\Delta lnEPU$  & -15.8287$^{***}$ & -15.7716$^{***}$ & -20.8516$^{***}$  &-20.7548$^{***}$  & 0.0724 & 0.0721\\
    $\Delta lnCPU$  & -9.0234$^{***}$  & -8.9968$^{***}$  & -51.9466$^{***}$
   & -51.6527$^{***}$   & 0.1100 & 0.0928\\
    $\Delta lnTPU$  & -18.0481$^{***}$ & -18.0336$^{***}$  &-23.5253$^{***}$
& -27.5122$^{***}$   & 0.4475$^{*}$ & 0.3034$^{***}$\\
    $\Delta lnGPR$  & -15.3510$^{***}$ & -15.3029$^{***}$  &-23.5435$^{***}$
   &-23.5737$^{***}$  & 0.1065 & 0.0833\\ 
    \bottomrule
    \end{tabular}
         \begin{tablenotes}
    \footnotesize
    \item Note: The unit root tests are performed on the log levels of the series. For ADF test \citep{Dickey-Fuller-1979-JASA, Dickey-Fuller-1981-Em}, the optimal lag length is chosen according to the smallest Schwarz information criterion (SIC). For both PP \citep{Phillips-Perron-1988-Bm} and KPSS \citep{Kwiatkowski-Phillips-Schmidt-Shin-1992-JEm} tests, the bandwidth is selected using the Newey-West Bartlett kernel. $\Delta$ refers to the first difference. The superscripts $^{***}$, $^{**}$, and $^{*}$ denote the statistical significance at the levels of 1\%, 5\%, and 10\%, respectively.
    \end{tablenotes}
    \end{threeparttable}
     }
  \label{Tab:UnitRootTest}
\end{table*}

\subsection{Results of NARDL models}
\label{S5.2:NARDL}

Following \cite{Shin-Yu-Greenwood-Nimmo-2014-Festschrift}, we use the Wald test to capture the long- and short-run asymmetric effects of the uncertainties on the total spillovers. The results shown in Table~\ref{Tab:AsymmetricTest} reveal significant short-run asymmetric effects of $ln CPU$ on $ln TSI_{\tau=0.95}$. For long-run asymmetric effects, $ln EPU$ is significant to $TSI_{\tau=0.5}$, and $ln CPU$, $ln TPU$, and $ln GPR$ are significant to $TSI_{\tau=0.05}$.

We select lag orders of NARDL models as (1, 1, 1, 1, 1, 0), (3, 1, 1, 1, 3, 0), and (1, 3, 3, 1, 1, 0) for $TSI_{\tau=0.5}$, $TSI_{\tau=0.05}$, and $TSI_{\tau=0.95}$, respectively. The $F_{PSS}$ statistics indicate significant long-run cointegration. Results are shown in Table~\ref{Tab:NARDL}. For $TSI_{\tau=0.5}$, contemporaneous $\Delta ln EPU^+$, $\Delta ln CPU^+$, and $\Delta ln TPU^-$ have significant impacts on $\Delta ln TSI_{\tau=0.5}$, with the coefficients are 0.17, -0.113, and -0.051, respectively. The long-run results demonstrate that $ln EPU^-$ and $COVID-19$ are positively related to $ln TSI_{\tau=0.5}$, while $ln TPU^+$ and $ln TPU^-$ are negatively related to $ln TSI_{\tau=0.5}$. When move to the results of $TSI_{\tau=0.05}$, we find that the coefficients of contemporaneous variables are not statistically significant. The first lag of $\Delta ln GPR^+$ and the second lag of $\Delta ln GPR^-$ have significant impact on $\Delta ln TSI_{\tau=0.05}$. Meanwhile, the long-run results show that the coefficient is significantly positive for $ln TPU^-$ and significantly negative for $ln GPR^+$ and $ln GPR^-$. For $TSI_{\tau=0.95}$, coefficients of contemporaneous and first lag $\Delta ln EPU^+$ are significant. As with $\Delta ln CPU$, the coefficients of the contemporaneous and all the lags of $\Delta ln CPU^+$ are negative, while they are all positive for the contemporaneous and all the lags of $\Delta ln CPU^-$. This is consistent with the asymmetric test in Table~\ref{Tab:AsymmetricTest}. Both $\Delta ln TPU^+$ and $\Delta ln TPU^-$ have significant negative coefficients, with similar magnitudes of 0.020 and 0.021. Additionally, panel C of Table~\ref{Tab:NARDL} reports the results of residual diagnostics tests, which approve the adequacy of the selected NARDL models.

\begin{table*}[!t]
  \centering
  \setlength{\abovecaptionskip}{0.1cm}
  \caption{Results of ARDL models.}
     \setlength{\tabcolsep}{4.3mm}{
       \begin{threeparttable}
    \begin{tabular}{lllllll}
    \toprule
      Variables & \multicolumn{2}{c}{$TSI_{\tau=0.5}$} & \multicolumn{2}{c}{$TSI_{\tau=0.05}$} & \multicolumn{2}{c}{$TSI_{\tau=0.95}$}\\
      \cmidrule(r){2-3} \cmidrule(r){4-5} \cmidrule(r){6-7}
       & Coeff.  & $p$-val. & Coeff. & $p$-val.& Coeff. & $p$-val.  \\
    \midrule
    \multicolumn{7}{l}{\textit{Panel A: Short-run results}}\\
    $Intercept$ & 0.705     & 0.000$^{***}$   & 1.699  &  0.000$^{***}$  & 0.871 & 0.000$^{***}$ \\
    $\Delta lnEPU_t$     & 0.051 & 0.410   &  0.048   &  0.092$^{*}$  & 0.074 & 0.003$^{***}$ \\
    $\Delta lnEPU_{t-1}$   &    &   &     &    & -0.070 & 0.009$^{***}$ \\
    $\Delta lnEPU_{t-2}$   &    &  &     &    & -0.047 & 0.063$^{*}$ \\
    $\Delta lnCPU_t$         & -0.036     & 0.312   &  -0.001    & 0.926   & -0.014 & 0.290\\
    $\Delta lnCPU_{t-1}$     & 0.096 &  0.010$^{**}$  &   &   & 	0.007 & 0.649\\
    $\Delta lnCPU_{t-2}$     & 0.056 &  0.135  &   &   & 	0.026 & 0.058$^{*}$\\
    $\Delta lnCPU_{t-3}$     & 0.112 &  0.002$^{***}$  &   &   & 	  &  \\
    $\Delta lnTPU_t$         & -0.015  & 0.302   & 0.001 & 0.950 & -0.020 & 0.000$^{***}$\\
    $\Delta lnTPU_{t-1}$     &    &    & 0.017  & 0.043$^{**}$ &  & \\
    $\Delta lnTPU_{t-2}$     & 	   &    & 0.015 & 0.037$^{**}$ &  & \\
    $\Delta lnGPR_t$     & 	-0.041  & 0.432   &  -0.051  & 0.040$^{**}$ & 0.028 & 0.158\\
    $\Delta lnGPR_{t-1}$     & 	-0.176 & 0.001$^{***}$  & 0.046  & 0.087$^{*}$ & -0.039 & 0.060$^{*}$\\
    $\Delta lnGPR_{t-2}$     &       &    & 0.051 & 0.041$^{**}$ &  & \\
    $ECT$  & 	-0.173     & 0.000$^{**}$   & -0.348  &  0.000$^{***}$  & -0.293 & 0.000$^{***}$\\
     \multicolumn{7}{l}{\textit{Panel B: Long-run results}}\\
    $Intercept$ & 4.075     & 0.031$^{**}$   & 4.873  &  0.000$^{***}$  & 2.973 & 0.000$^{***}$ \\
    $lnEPU$     & -0.257    & 0.485   & 0.137  &  0.083$^{*}$  & 0.267 & 0.009$^{***}$ \\
    $lnCPU$     & -0.548    & 0.059$^{*}$   &  -0.005  & 0.928   & -0.122 & 0.121\\
    $lnTPU$     & -0.042     & 0.486   &  -0.024  & 0.100  & -0.037  & 0.009$^{***}$\\
    $lnGPR$     & 0.712    & 0.051$^{*}$   &   -0.205 &  0.015$^{**}$ & 0.185 & 0.019$^{**}$\\
    $COVID-19$  & 0.969   & 0.001$^{***}$   &   -0.015 &  0.796 & -0.007 & 0.902\\ 
    \multicolumn{7}{l}{\textit{Panel C: Diagnostics tests}}\\
    $F_{PSS}$  &  3.382     & 0.096$^{*}$     &  4.265 &  0.023$^{**}$  & 4.252 &  0.024$^{**}$\\
    BG   & 0.007  & 0.933  &  1.037 & 0.308 &  0.088 & 0.767\\
    BP    & 28.893  & 0.011$^{**}$   & 15.562 & 0.341 & 20.828 & 0.142\\
    Ramsey RESET    & 2.335  & 0.009$^{***}$   & 1.037 & 0.427 & 0.998 & 0.467\\
    CUSUM   & 0.486 & 0.659   &  0.619   &  0.377 &  0.643 & 0.335\\
    \bottomrule
    \end{tabular}
         \begin{tablenotes}
    \footnotesize
    \item  The superscripts $^{***}$, $^{**}$, and $^{*}$ denote the statistical significance at the levels of 1\%, 5\%, and 10\%, respectively.
    \end{tablenotes}
    \end{threeparttable}
     }
  \label{Tab:ARDL}
\end{table*}

    \begin{table*}[!t]
  \centering
  \setlength{\abovecaptionskip}{0.1cm}
  \caption{Results of the Wald test for asymmetric effects.}
     \setlength{\tabcolsep}{6mm}{
       \begin{threeparttable}
    \begin{tabular}{lllllll}
    \toprule
      Variables & \multicolumn{2}{c}{$TSI_{\tau=0.5}$} & \multicolumn{2}{c}{$TSI_{\tau=0.05}$} & \multicolumn{2}{c}{$TSI_{\tau=0.95}$}\\
      \cmidrule(r){2-3} \cmidrule(r){4-5} \cmidrule(r){6-7}
       & Coeff.  & $p$-val. & Coeff. & $p$-val.& Coeff. & $p$-val.  \\
    \midrule
    \multicolumn{7}{l}{\textit{Panel A: Short-run results}}\\
     $W_{EPU}$      & 0.010     & 0.922 &   0.855 & 0.358 & 0.508 & 0.478\\
    $W_{CPU}$      & 1.219     & 0.273 &   0.172 & 0.679 & 4.384 & 0.040$^{**}$\\
    $W_{TPU}$      & 0.438     & 0.510 &  0.201 & 0.655 & 0.005  & 0.942 \\
    $W_{GPR}$      & 1.507     & 0.223 &  0.047 & 0.830 & 0.455  & 0.502 \\
     \multicolumn{7}{l}{\textit{Panel B: Long-run results}}\\
    $W_{EPU}$      & 4.360      & 0.016$^{**}$ &   1.638 & 0.201 & 2.954 & 0.058$^{*}$\\
    $W_{CPU}$      & 0.920     & 0.402 &    2.937 & 0.059$^{*}$ & 1.728  & 0.184\\
    $W_{TPU}$      & 2.209     & 0.116 &  5.291 & 0.007$^{***}$ & 1.798  & 0.172 \\
    $W_{GPR}$      & 0.131     & 0.877 &  4.795 & 0.011$^{**}$ & 0.656  & 0.522 \\
     \bottomrule
       \end{tabular}
         \begin{tablenotes}
    \footnotesize
    \item  The superscripts $^{***}$, $^{**}$, and $^{*}$ denote the statistical significance at the levels of 1\%, 5\%, and 10\%, respectively.
    \end{tablenotes}
    \end{threeparttable}
     }
  \label{Tab:AsymmetricTest}
  \end{table*}

    \begin{table*}[!t]
  \centering
  \setlength{\abovecaptionskip}{0.1cm}
  \caption{Results of NARDL models.}
     \setlength{\tabcolsep}{6mm}
       \begin{threeparttable}
    \footnotesize{
    \begin{tabular}{lllllll}
    \toprule
      Variables & \multicolumn{2}{c}{$TSI_{\tau=0.5}$} & \multicolumn{2}{c}{$TSI_{\tau=0.05}$} & \multicolumn{2}{c}{$TSI_{\tau=0.95}$}\\
      \cmidrule(r){2-3} \cmidrule(r){4-5} \cmidrule(r){6-7}
       & Coeff.  & $p$-val. & Coeff. & $p$-val.& Coeff. & $p$-val.  \\
    \midrule
    \multicolumn{7}{l}{\textit{Panel A: Short-run results}}\\
    $Inetrcept$ & 1.507  & 0.000$^{***}$ &  2.354 & 0.000$^{***}$  & 1.700 & 0.000$^{***}$\\
    $\Delta lnTSI_{t-1}$ &   &  & 0.036  & 0.700 &  & \\
    $\Delta lnTSI_{t-2}$ &   &  & 0.170 & 0.036$^{**}$ &  & \\
    $\Delta lnEPU_t^+$     &  0.170 & 0.080$^{*}$ & -0.002 & 0.958 & 0.132 & 0.001$^{***}$ \\
    $\Delta lnEPU_{t-1}^+$ &   &  &    &   & -0.088 & 0.027$^{**}$\\
    $\Delta lnEPU_{t-2}^+$   &   &  &    &   & -0.049 & 0.211\\
    $\Delta lnEPU_t^-$     &  0.149 & 0.205 & 0.081 & 0.110 & 0.047 & 0.340 \\
    $\Delta lnEPU_{t-1}^-$ &   &  &   &   & -0.034 & 0.470\\
    $\Delta lnEPU_{t-2}^-$   &   &  &   &   & -0.010 & 0.836\\
    $\Delta lnCPU_t^+$     & -0.113 & 0.049$^{**}$ & 0.006 & 0.831 & -0.059 & 0.019$^{**}$\\
    $\Delta lnCPU_{t-1}^+$ &   &  &  &  & -0.029 & 0.237\\
    $\Delta lnCPU_{t-2}^+$   &   &  &   &   & -0.011 & 0.633\\
    $\Delta lnCPU_t^-$     & 0.010 & 0.865 & -0.015 & 0.556 & 0.017 & 0.477\\
    $\Delta lnCPU_{t-1}^-$ &   &  &   &   & 0.054 & 0.049$^{**}$\\
    $\Delta lnCPU_{t-2}^-$   &   &  &   &   & 0.064 & 0.028$^{**}$\\
    $\Delta lnTPU_t^+$       & -0.017 & 0.499 & -0.007 & 0.546 & -0.021 & 0.034$^{**}$ \\
    $\Delta lnTPU_t^-$       & -0.051 & 0.074$^{*}$ & 0.003 & 0.779 & -0.020 & 0.091$^{*}$\\
    $\Delta lnGPR_t^+$     & -0.126 &  0.118 & -0.029 & 0.395 & -0.001 & 0.989\\
    $\Delta lnGPR_{t-1}^+$ &   &  & 0.128  & 0.002$^{***}$ &  & \\
    $\Delta lnGPR_{t-2}^+$   &  &  & 0.031 & 0.416 &  & \\
    $\Delta lnGPR_t^-$     & 0.102 & 0.311 & -0.038 & 0.395 & 0.050 & 0.194\\
    $\Delta lnGPR_{t-1}^-$ &   &  & 0.019 & 0.688 &  & \\
    $\Delta lnGPR_{t-2}^-$   &  &  & 0.117 & 0.005$^{***}$ &  & \\
    $ECT$  &  -0.392   &  0.000$^{***}$  & -0.527  & 0.000$^{***}$   & -0.375 & 0.000$^{***}$ \\
     \multicolumn{7}{l}{\textit{Panel B: Long-run results}}\\
    $Intercept$ & 3.839 & 0.000$^{***}$   & 4.466  &  0.000$^{***}$  & 4.537  &  0.000$^{***}$ \\ 
    $lnEPU^+$     & 0.186     & 0.441   & 0.001  & 0.997  & 0.293 & 0.042$^{**}$ \\
    $lnEPU^-$     & 0.634     & 0.007$^{***}$   & 0.096  &  0.220  & 0.265 & 0.048$^{**}$ \\
    $lnCPU^+$      & 0.094     & 0.447   &  0.057  & 0.141   & -0.124 & 0.109\\
    $lnCPU^-$      & -0.025     & 0.872   &  -0.016  & 0.743   & -0.151 & 0.102\\
    $lnTPU^+$       & -0.122     & 0.017$^{**}$   &  -0.012  & 0.451  & -0.038  & 0.095$^{*}$\\
    $lnTPU^-$       & -0.120     & 0.014$^{**}$   &  0.027  & 0.080$^{*}$  & -0.039  & 0.068$^{*}$\\
    $lnGPR^+$      & 0.082     & 0.621   &   -0.124 &  0.078$^{*}$ & 0.031 & 0.704\\
    $lnGPR^-$      & 0.027     & 0.890   &   -0.232 &  0.005$^{***}$ & 0.102 & 0.293\\
    $COVID-19$    & 0.402     & 0.003$^{***}$   &  -0.018 &  0.703 & -0.063 & 0.256\\ 
    \multicolumn{7}{l}{\textit{Panel C: Diagnostics tests}}\\
    $F_{PSS}$  & 3.218 & 0.062$^{*}$ & 3.569 & 0.027$^{**}$  & 3.044 &  0.091$^{*}$\\
    BG   & 0.841   &  0.359  & 0.798   & 0.372 & 0.123 & 0.726\\
    BP     & 31.849 & 0.023$^{**}$ & 28.498 & 0.240 & 30.123 & 0.263\\
    Ramsey RESET    & 0.718 & 0.781   & 0.827 & 0.688 & 0.885 & 0.624 \\
    CUSUM   &  0.373 &  0.891 &  0.680 &  0.278 & 1.011 & 0.031$^{**}$ \\
    \bottomrule
       \end{tabular}
    }
         \begin{tablenotes}
    \footnotesize
    \item  The superscripts $^{***}$, $^{**}$, and $^{*}$ denote the statistical significance at the levels of 1\%, 5\%, and 10\%, respectively.
    \end{tablenotes}
    \end{threeparttable}
     
  \label{Tab:NARDL}
  \end{table*}

\section{Conclusion and policy implications}
\label{S6:Conclude}

This paper use a quantile regression-based \cite{Diebold-Yilmaz-2012-IntJForecast,Diebold-Yilmaz-2014-JE} spillover measure to explore the return connectedness between food, fossil energy, and clean energy markets at the median and extreme quantiles. Additionally, we examine the role of external uncertainties on the spillover effects under different market conditions.

 Our empirical results show that the return connectedness between these markets is much stronger at the tails (61.47\% for left tail and 57.91\% for right tail) than at the median (23.02\%). The total spillover index presents a U-shaped curve across quantiles, indicating that returns between these markets are more tightly connected during the extreme market conditions. The net spillover analysis reveals that fossil energy market always act as the net receiver, while clean energy market primarily serves as the net transmitter. The dynamic analysis shows that spillover effects vary over time and intensify during period of extreme events, such as the signing and implementation of the Paris Agreement, and the COVID-19 pandemic. Furthermore, results from the ARDL and NARDL models show that external uncertainties have statistically significant impacts on total spillovers. At the median quantile, CPU, GPR, and the COVID-19 pandemic are the important drivers of spillovers. At the extreme quantiles, EPU, TPU, and GPR act as main drivers. In addition, the results of NARDL models reveal the asymmetric effects of external uncertainties.

Our findings have several practical implications for cross-market investments in food and energy markets. First, the significant return spillovers, particularly under extreme market conditions, highlight the risk contagions between these markets. Investors should carefully monitor these risks and implement strategies to manage cross-market exposures. Second, as fossil energy primarily acts as a net receiver of shocks, investors in fossil energy markets need to track developments in food and clean energy markets and diversify their portfolios by incorporating food and clean energy assets. Third, given the significant influence of external uncertainties, investors should adjust their strategies during periods of heightened uncertainty related to EPU, CPU, TPU, or GPR to mitigate potential risks.

The results also carry critical implications for policymakers. First, the transition from fossil energy to clean energy is an important issue, which requires well-designed policies that account for the interconnectedness between these markets under varying conditions. Second, our research reveals the significant impact of external uncertainties on the connectedness between food and energy markets. Therefore, policymakers should closely monitor the changes in external uncertainties and employ useful policy tools to achieve policy coordination when uncertainty shocks occur, which is of great importance to ensuring the stability of food and energy markets.

\section*{Acknowledgment}

This work was supported by the Excellent Youth Project of Hunan Provincial Department of Education (Grant Number: 23B0425) and the Youth Project of Hunan Provincial Social Science Fund (Grant Number: 23YBQ080)



%
\bibliography{Bib1,Bib2}

\end{document}